\documentclass[a4paper,11pt]{article}
\usepackage[textwidth=16.2cm,textheight=22cm]{geometry}
\usepackage{amsfonts}
\usepackage{amssymb}
\usepackage{amsmath}
\usepackage{amscd}
\usepackage{latexsym}
\usepackage{bbm}
\usepackage{epic}
\usepackage{pst-node}

\usepackage[latin1]{inputenc}

\catcode`@=11
\renewcommand{\section}{\@startsection{section}{1}{0pt}{\medskipamount}
{\medskipamount}{\large\bf}}
\numberwithin{equation}{section}
\catcode`@=12

\def\del{\partial}
\def\a{\alpha}
\def\b{\beta}
\def\g{\gamma}

\def\eps{\epsilon}
\def\ve{\varepsilon}

\def\l{\lambda}
\def\m{\mu}
\def\n{\nu}
\def\r{\rho}

\def\s{\sigma}
\def\t{\tau}

\def\la{\lambda}

\newcommand{\ZR}{\mathbbm R}
\newcommand{\ZZ}{\mathbbm Z}

\newcommand{\CA}{\mathcal{A}}       

\newcommand{\CP}{\mathcal{P}}       
\newcommand{\CN}{\mathcal{N}}      
\newcommand{\CM}{\mathcal{M}}       
\newcommand{\CO}{\mathcal{O}}      
\newcommand{\CR}{\mathcal{R}}

\def\_{\;\;}
\def\^{\wedge}

\def\pd{\mbox{$\partial$}}
\def\diff{\mbox{d}}

\def\Tr{{\rm Tr}}
\def\eqn#1{eq.~(\ref{#1})}

\def\sfrac#1#2{{\textstyle\frac{#1}{#2}}}
\def\>{\rangle}
\def\<{\langle}
\def\+{\dagger}
\def\={\ =\ }

\def\open{{\rm open}}
\def\closed{{\rm closed}}

\def\SYM{{\rm SYM}}
\def\SUGRA{{\rm SUGRA}}
\def\and{\qquad\textrm{and}\qquad}

\begin{document}

\begin{titlepage}
\renewcommand{\thefootnote}{\fnsymbol{footnote}}
SLAC--PUB--13848 \hfill AEI-2009-110
\hfill ITP-UH-18/09 \hfill arXiv:0911.5704
\vskip 2cm
\centerline{\Large{\bf {$\CR^4$ counterterm and $E_{7(7)}$
symmetry in maximal supergravity}}}
\vskip 1.0cm
\centerline{Johannes Br\"odel${}^{a,b}$ and  Lance J. Dixon${}^{c}$
}
\vskip .5cm
\centerline{${}^a$ \it {Max-Planck-Institut f\"{u}r
Gravitationsphysik}}
\centerline{\it {Albert-Einstein-Institut, Golm, Germany}}
\vskip .5cm
\centerline{${}^b$ \it {Institut f\"ur Theoretische Physik}}
\centerline{\it {Leibniz Universit\"at Hannover, Germany}}
\vskip .5cm
\centerline{${}^c$ \it {SLAC National Accelerator Laboratory}}
\centerline{\it {Stanford University}}
\centerline{\it {Stanford, CA 94309, USA}}
\vskip 1.5cm
\centerline{\bf {Abstract}}
\vskip .5cm
\noindent 
The coefficient of a potential $\CR^4$ counterterm in $\CN=8$
supergravity has been shown previously to vanish in an explicit three-loop
calculation.  The $\CR^4$ term respects $\CN=8$ supersymmetry; hence
this result poses the question of whether another symmetry could be
responsible for the cancellation of the three-loop divergence. 
In this article we investigate possible restrictions from the 
continuous 
coset 
symmetry $E_{7(7)}/SU(8)$, exploring the limits as a single scalar becomes
soft, as well as a double-soft scalar limit relation
derived recently by Arkani-Hamed {\it et al.}\ \  
We implement these relations for the matrix elements of
the $\CR^4$ term that occurs in the low-energy expansion of closed-string
tree-level amplitudes.  We find that the matrix elements of $\CR^4$ that
we investigated all obey the double-soft scalar limit relation, 
including certain non-maximally-helicity-violating six-point
amplitudes.  However, the single-soft limit does not vanish for this
latter set of amplitudes, which suggests that the $E_{7(7)}$ symmetry
is broken by the $\CR^4$ term.
\vfill
\end{titlepage}

\tableofcontents
\newpage


\section{Introduction}

Divergences of four-dimensional gravity theories have been under
investigation practically since the advent of quantum field theory.
While pure gravity
can be shown to be free of ultraviolet divergences at one loop, the
addition of scalars or other particles renders the theory
nonrenormalizable~\cite{tHooft1974bx}. At the two-loop level,
the counterterm
\begin{equation}
R^3\ \equiv\ R_{\m\n}^{\l\r}\,R^{\s\t}_{\la\r}\,R^{\m\n}_{\s\t}
\end{equation}
has been shown to respect all symmetries, to exist
on-shell~\cite{Kallosh1974yh,vanNieuwenhuizen1976vb}
and to have a nonzero coefficient for pure
gravity~\cite{Goroff1985th,vandeVen1991gw}.

Supersymmetry is known to improve the ultraviolet behavior of many
quantum field theories.  In fact, supersymmetry forbids the $R^3$ 
counterterm in any supersymmetric version of four-dimensional gravity,
provided that all particles are in the same multiplet as the
graviton~\cite{Grisaru1976nn,Deser1977nt,Tomboulis1977wd}.
That is because the operator $R^3$ generates a scattering amplitude
that can be shown to vanish by supersymmetric Ward identities
(SWI)~\cite{Grisaru1976nn,Grisaru1976vm,Grisaru1977px,%
Parke1985pn,Kunszt1985mg}.
However, the next possible
counterterm~\cite{Deser1977nt,Ferrara1977mv,Deser1978br,%
Howe1980th,Kallosh1980fi} is
\begin{equation}\label{eqn:R4}
R^4
\ \equiv\ t_8^{\m_1\n_1\ldots\m_4\n_4} t_8^{\r_1\s_1\ldots\r_4\s_4}
R_{\m_1\n_1\r_1\s_1} R_{\m_2\n_2\r_2\s_2}
R_{\m_3\n_3\r_3\s_3} R_{\m_4\n_4\r_4\s_4} \,,
\end{equation}
where $t_8$ is defined in eq.~(4.A.21) of ref.~\cite{Schwarz1982jn}.
This operator, also known as the square of the Bel-Robinson
tensor~\cite{Bel1958}, on dimensional grounds can appear as a counterterm
at three loops.  It is compatible with supersymmetry, not just
${\cal N}=1$ but all the way up to maximal $\CN=8$ supersymmetry.
This property follows from the appearance of $R^4$ in the 
low-energy effective action of the $\CN=8$ supersymmetric
closed superstring~\cite{Gross1986iv}; indeed, it represents 
the first correction term beyond the limit of
$\CN=8$ supergravity~\cite{Green1982sw},
appearing at order $\a^{\prime3}$.  We denote by $\CR^4$ the $\CN=8$
supersymmetric multiplet of operators containing $R^4$.

We note that beyond the four-point level, and in more than four
dimensions, it is possible to distinguish at least
one other quartic combination of Riemann tensors, maintaining
${\cal N}=8$ supersymmetry.  In the notation of
refs.~\cite{Peeters2000qj,Howe2004pn}, the $R^4$ term appearing in
the tree-level closed superstring effective action in ten dimensions
is actually $e^{-2\phi}(t_8t_8 - {1\over8} \eps_{10} \eps_{10}) R^4$,
where $\eps_{10}$ is the ten-dimensional totally antisymmetric tensor,
and $\phi$ is the (ten-dimensional) dilaton. The dilaton is also the 
string loop-counting parameter, so that terms in the effective
action at $L$ loops are proportional to $\exp(-2(1-L)\phi)$ (in string
frame).  The corresponding term in the one-loop effective action in the
IIA string theory differs from the IIB case in the sign of
the $\eps_{10} \eps_{10}$ term, and is proportional to
$(t_8t_8 + {1\over8} \eps_{10} \eps_{10}) R^4$.  In four dimensions, 
the $\eps_{10} \eps_{10}$ terms vanish.  However, the different
possible dependences of $R^4$ terms on the dilaton persist, and become
more complicated, because the dilaton resides in the 70 scalars
of $\CN=8$ supergravity, which are members of the $\mathbf{70}$
representation of $SU(8)$, and the $R^4$ prefactor should be $SU(8)$
invariant.  Green and Sethi~\cite{Green1998by} found powerful constraints
on the possible dependences in ten dimensions using supersymmetry alone;
indeed, only tree-level ($e^{-2\phi}$) and one-loop (constant)
terms are allowed.  It would be very interesting to examine the analogous
supersymmetry constraints in four dimensions.

The issue of possible counterterms in maximal $\CN=8$
supergravity~\cite{Cremmer1978ds,Cremmer1979up}
is under perpetual investigation. Many of the current arguments rely on
(linearized) superspace formulations and nonrenormalization 
theorems~\cite{Howe1983sr,Howe2002ui}, which in turn depend
on the existence of an off-shell superspace formulation.  It was
a common belief for some time that a superspace formulation of
maximally-extended supersymmetric theories could be achieved employing
off-shell formulations with at most half of the supersymmetry
realized. On the other hand, 
an off-shell harmonic superspace with $\CN=3$ supersymmetry
for $\CN=4$ super-Yang-Mills (SYM) theory was constructed
a while ago~\cite{Galperin1984bu}.
Assuming the existence of a similar description realizing
six of the eight supersymmetries of $\CN=8$ supergravity would postpone the
onset of possible counterterms at least to the five-loop level, 
while realizing seven of eight would postpone it to the six-loop
level~\cite{Howe2002ui}.  However, an explicit construction of such
superspace formalisms has not yet been achieved in the gravitational case.

Another way to explore the divergence structure of $\CN=8$
supergravity is through direct computation of on-shell multi-loop
graviton scattering amplitudes.  The two-loop four-graviton scattering
amplitude~\cite{Bern1998ug} provided the first hints that the
$\CR^4$ counterterm might have a vanishing coefficient at three loops.
The full three-loop computation then demonstrated this vanishing
explicitly~\cite{Bern2007hh,Bern2008pv}.  A similar cancellation
has been confirmed at four loops recently~\cite{Bern2009kd}.
The latter cancellation in four dimensions is not so surprising for
the four-point amplitude, because operators of the form $\del^2R^4$
can be eliminated in favor of $R^5$ using equations of
motion~\cite{Richards2008jg}, and it has been shown that there
is no $\CN=8$ supersymmetric completion of
$R^5$~\cite{Kallosh2009jb,Drummond2003ex}.
(This is consistent with the absence of $R^5$ terms from the 
closed-superstring effective action~\cite{Stieberger2009rr}.)
On the other hand, the explicit multi-loop amplitudes show an
even-better-than-finite ultraviolet behavior, as good as that for
${\cal N}=4$ super-Yang-Mills theory, which 
strengthens the evidence for a yet-unexplored underlying symmetry
structure.

There are also string- and M-theoretic arguments for the excellent
ultraviolet behavior observed to date.  Using a nonrenormalization
theorem developed in the pure spinor formalism for the closed
superstring~\cite{Berkovits}, Green, Russo and Vanhove
argued~\cite{GreenII} that the first divergence in $\CN=8$
supergravity might be delayed until nine loops.
(On the other hand, a very recent analysis of dualities and volume-dependence
in compactified string theory by the same authors~\cite{Green2010sp}
indicates a divergence at seven loops, in conflict with the previous
argument.)  Arguments based on M-theory dualities suggest the possibility of
finiteness to all loop
orders~\cite{DualityArgumentsI,DualityArgumentsII}.
However, the applicability of arguments based on string and M theory to
$\CN=8$ supergravity is subject to issues related
to the decoupling of massive states~\cite{GOS}.

There have also been a variety of attempts to understand the ultraviolet
behavior of $\CN=8$ supergravity more directly at the amplitude level.
The ``no triangle'' hypothesis~\cite{Bern1998sv,MoreNoTriangle},
now a theorem~\cite{VanhoveNoTriangle,ArkaniHamed2008gz}, states in essence
that the ultraviolet behavior of $\CN=8$ supergravity at one loop is as
good as that of $\CN=4$ super-Yang-Mills theory. It also implies many,
though not all, of the cancellations seen at higher loops~\cite{Finite}.
Some of the one-loop cancellations are not just due to supersymmetry,
but to other properties of gravitational theories~\cite{UnexpectedOneloop},
including their non-color-ordered nature~\cite{VanhoveOrderless}.

These one-loop considerations, and the work of ref.~\cite{Howe2002ui},
suggest that conventional $\CN=8$ supersymmetry alone may not be 
enough to dictate the finiteness of $\CN=8$ supergravity.
However, since the construction of $\CN=8$
supergravity~\cite{CremmerJuliaScherk,Cremmer1978ds,Cremmer1979up}
it has been realized that another symmetry plays a key role ---
the exceptional, noncompact continuous symmetry $E_{7(7)}(\ZR)$,
or $E_{7(7)}$ for short.
Could this symmetry contribute somehow to an explanation of the
(conjectured) finiteness of the theory?

The general role of the $E_{7(7)}$ symmetry, regarding the finiteness of
maximal supergravity, has been a topic of constant discussion.
(Aspects of its action on the Lagrangian in light-cone gauge~\cite{E7Recent},
and covariantly~\cite{Kallosh2008ic,Hillmann2009zf},
have also been considered recently.)
A seven-loop $\CN=8$ supersymmetric counterterm was constructed in the past
by Howe and Lindstr\"om~\cite{Howe1980th}.  Although this counterterm
does not appear to be invariant under the nonlinear $E_{7(7)}$
symmetry~\cite{Kallosh1980fi}, 
the volume form for the on-shell $\CN=8$ superspace represents a second, 
$E_{7(7)}$-invariant, seven-loop counterterm ---
if it is nonvanishing~\cite{Bossard2009mn}.
Also, a manifestly $E_{7(7)}$-invariant counterterm was presented long ago
at eight loops~\cite{Howe1980th,Kallosh1981vz}; 
however, newer results using the light-cone
formalism cast a different light on the question~\cite{Kallosh2009db}.

In this article we investigate whether restrictions on the
appearance of the $\CR^4$ term could originate directly from the exceptional
symmetry.  One way to test whether $\CR^4$ is invariant under $E_{7(7)}$
is to utilize properties of the on-shell amplitudes that $\CR^4$ produces.
This method is convenient because it turns out that the amplitudes
can be computed, using string theory, even when a full nonlinear 
expression for $\CR^4$ in four dimensions is unavailable.
However, it is limited to the matrix elements produced by
the $\CR^4$ term in the tree-level string effective action.
As discussed earlier, there may be other possible ${\cal N}=8$
supersymmetric $\CR^4$ terms, distinguished for example by their
precise dependence on the scalar fields in the theory, which we will
not be able to probe in this way.

Arkani-Hamed, Cachazo and Kaplan (ACK)~\cite{ArkaniHamed2008gz}
provided a very useful tool for an amplitude-based approach.  Working
in pure $\CN=8$ supergravity, they showed recursively how generic 
amplitudes with one soft scalar particle vanish as the soft
momentum approaches zero.  This vanishing was first observed by Bianchi,
Elvang and Freedman~\cite{Bianchi2008pu}, and associated with the fact
that the scalars parametrize the coset manifold $E_{7(7)}/SU(8)$ and obey
relations similar to soft pion theorems~\cite{Adler1964um,Coleman}.
On the other hand, in the case of soft pion emission, the amplitude can
remain nonvanishing as the (massless) pion momentum vanishes,
due to graphs in which the pion is emitted off an external line;
a divergence in the adjacent propagator cancels a power of pion momentum
in the numerator from the derivative interaction.  In the supergravity case,
it was found that the external scalar emission graphs actually
vanish on-shell in the soft limit~\cite{Bianchi2008pu}.

ACK further considered in detail the emission of two
additional soft scalar particles from a hard scattering amplitude, and
thereby derived a relation between amplitudes differing by two in the
number of legs. The relation should hold for any theory with
$E_{7(7)}$ symmetry. If one could show agreement of the single-soft
limit and the ACK relation
for all amplitudes derived from a modified $\CN=8$ supergravity
action, in this case perturbing it by the $\CR^4$ term, then this
action should obtain no restrictions from $E_{7(7)}$.

Actually, for this conclusion to hold, $E_{7(7)}$ should remain
a good symmetry at the quantum level.  Although there is evidence in favor
of this, we know of no all-orders proof. At one loop, the cancellation
of anomalies for currents from the $SU(8)$ subgroup of $E_{7(7)}$
was demonstrated quite a while ago~\cite{Marcus1985yy}.  The analysis was
subtle because a Lagrangian for the vector particles cannot be written in
a manifestly $SU(8)$-covariant fashion.  Thus the vectors contribute to 
anomalies, cancelling the more-standard contributions from the fermions.
More recently, the question of whether the full $E_{7(7)}$ is a good
quantum symmetry has been re-examined using the methods of ACK.
He and Zhu recently showed that the infrared-finite part of single-soft 
scalar emission vanishes at one loop for an arbitrary number of 
external legs~\cite{He2008pb} as it does at tree
level.  (Earlier, Kallosh, Lee and Rube~\cite{Kallosh2008ru}
showed the vanishing of the four-point one-loop amplitude in the 
single-soft limit for complex momenta.) 
A similar argument by Kaplan~\cite{Kaplanprivate} shows that 
the double-soft scalar limit relation in $\CN=8$ supergravity can also 
be extended to one loop.  These results support the conjecture that
the full continuous $E_{7(7)}(\ZR)$ is a good quantum symmetry of the 
theory, at least at the one-loop level.  Beyond perturbation theory,
assuming that black holes contribute to graviton scattering amplitudes,
there is good reason to believe that the continuous symmetry will be broken,
but that a discrete subgroup $E_{7(7)}(\ZZ)$ will survive.  However,
non-perturbative considerations are far beyond the scope of this article.

The purpose of this article is to test the $E_{7(7)}$ invariance 
of \eqn{eqn:R4}, by exploring the validity of the single-soft limits
and the ACK relation for the four-dimensional $\CN=8$ supergravity action,
modified by adding the supersymmetric extension of the $R^4$ term that
appears in the tree-level closed superstring effective action. 
The bulk of the article is devoted to the construction of amplitudes
produced by this term.
As we will see, we need to go to six-point
next-to-maximal-helicity-violating (NMHV) amplitudes 
to get the first nontrivial result. The strategy for obtaining information
about higher-order $\a'$-terms in closed-string scattering is the same as 
used in a recent article by Stieberger~\cite{Stieberger2009rr}: We will fall 
back to open-string calculations~\cite{Stieberger2007am} and 
derive the corresponding closed-string results by employing the
Kawai-Lewellen-Tye (KLT)~\cite{KLT} relations. 

The remainder of this article is organized as follows.
Sections \ref{sec:coset} and \ref{sec:corr} collect the background
information on symmetries of $\CN=8$ supergravity, including
the double-soft scalar limit of amplitudes, and they illuminate the
state and availability of open-string amplitude calculations.
In section \ref{sec:settingupcalc} the calculation is set up. We start
by introducing the KLT relations connecting open- and closed-string
amplitudes in subsection \ref{sec:KLT}.  A suitable amplitude for probing
the double-soft scalar limit relation is singled out in
subsection~\ref{sec:suitamp}.  The $\CN=1$ supersymmetric Ward identities 
needed to make use of the available open-string amplitudes are described
in detail in subsections~\ref{sec:SWI}, \ref{sec:N1SWI}
and~\ref{sec:seconddiamond}.  The main result of this article, the 
testing of possible restrictions originating from $E_ {7(7)}$ symmetry,
by employing the single- and double-soft scalar limit relations
on amplitudes produced by the $\CR^4$ term, is presented in
section~\ref{sec:calculation}.
In section~\ref{sec:conclusion} we draw our conclusions.


\section{Coset structure, hidden symmetry and double-soft
limit}\label{sec:coset}

The physical field content of the maximal supersymmetric gravitational
theory in four dimensions, $\CN=8$
supergravity~\cite{Cremmer1978ds,Cremmer1979up},
consists of a vierbein (or graviton), 8 gravitini, 28 abelian
gauge fields, 56 Majorana gauginos of either helicity, and 70 real
(or 35 complex) scalars, which can be collected together in a single
massless $\CN=8$ (on-shell) supermultiplet.

Starting from the fact that the vector bosons form an antisymmetric
tensor representation of $SO(8)$ in the ungauged theory,
Bianchi identities and equations of
motion can be considered in order to realize a much larger symmetry, which
leads to the notion of generalized electric-magnetic 
duality transformations. Investigating
these transformations more closely and enlarging the corresponding duality
group maximally by adding further scalars, not all of which turn out to be
physical.  After gauging a resulting local $SU(8)$ symmetry in order to
reduce the degrees of freedom of the generalized duality group, $70$
physical scalars remain.  These scalars parameterize the coset
$\sfrac{E_{7(7)}}{SU(8)}$~\cite{Cremmer1979up,deWit1982ig}, where
$E_{7(7)}$ denotes a noncompact real form of $E_7$, which has $SU(8)$ as
its maximal compact subgroup.  In other words, the scalars
can be identified with the noncompact generators of $E_{7(7)}$. 
The resulting gauge is called \textit{unitary}.

More explicitly, in unitary gauge the $63$ compact generators
$T_I^J$ of $SU(8)$ can be joined with 70 generators $X_{I_1\ldots I_4}$
to form the adjoint representation of $E_{7(7)}$.  Here $X_{I_1\ldots I_4}$
transforms under $SU(8)$ in the
four-index antisymmetric tensor representation $(I,J=1,\ldots, 8)$. The
commutation relations between those generators are given schematically by
\begin{equation}\label{eqn:commrel}
 [T,T]\sim T\,,\quad[X,T]\sim X\,, \and[X,X]\sim T\,.
\end{equation}
The first commutator is just the usual $SU(8)$ Lie algebra, and the 
second one follows straightforwardly from the identification of $X$
with the $\mathbf{70}$ of $SU(8)$.  The more nontrivial statement about
$E_{7(7)}$ invariance resides in the third commutator in \eqn{eqn:commrel}.
Assuming the two scalars to be represented as $X_1^{I_1\ldots I_4}$
and $X_{2\,I_5\ldots I_8}$, where the upper-index version
can be obtained by employing the $SU(8)$-invariant tensor, 
\begin{equation}
X^{I_1\ldots I_4}\ =\ \frac{1}{24}
\ve^{I_1I_2I_3I_4I_5I_6I_7I_8}X_{I_5\ldots I_8}\,,
\end{equation}
the third relation reads explicitly
(see {\it e.g.} ref.~\cite{ArkaniHamed2008gz}),
\begin{equation}\label{eqn:Cartan}
-i \, [X_1^{I_1\ldots I_4},X_{2\,I_5\ldots I_8}]
\ =\ 
\ve^{JI_2I_3I_4}_{I_5I_6I_7I_8}T^{I_1}_J
\,+\,\ve^{I_1JI_3I_4}_{I_5I_6I_7I_8}T^{I_2}_J
\,+\,\ldots\,+\,\ve^{I_1I_2I_3I_4}_{I_5I_6I_7J}T^{J}_{I_8}\,.
\end{equation}
Here $\ve^{I_1I_2I_3I_4}_{I_5I_6I_7I_8}=1,-1,0$ if the
upper index set is an even, odd or no permutation of the lower set,
respectively.  (For a more general discussion of the properties 
of $E_{7(7)}$, see appendix B of ref.~\cite{Cremmer1979up}.)

Amplitudes in $\CN=8$ supergravity are invariant under $SU(8)$ rotations
by construction. On the other hand, the action of the coset symmetry
$\frac{E_{7(7)}}{SU(8)}$ on amplitudes is not obvious. One can understand
the connection by recalling that the vacuum state of the theory is specified
by the expectation values of the physical scalars. Because the scalars are
Goldstone bosons, the soft emission of scalars in an amplitude changes
the expectation value and moves the theory to another point in the
vacuum manifold.

Arkani-Hamed, Cachazo and Kaplan~\cite{ArkaniHamed2008gz} used
the BCFW recursion relations~\cite{Britto2004ap,Britto2005fq} to
investigate how the noncompact part of $E_{7(7)}$ symmetry controls
the soft emission of scalars in $\CN=8$ supergravity. Consider
first the emission of a single soft scalar (which was also studied in
refs.~\cite{Bianchi2008pu,Kallosh2008rr}).  The corresponding amplitudes
can be traced back via the BCFW recursion relations to the three-particle
amplitude, whose vanishing in the soft limit can be shown explicitly.
Hence the emission of a single scalar from any amplitude vanishes in
$\CN=8$ supergravity,
\begin{equation}\label{eqn:singlesoft}
M_{n+1}(1,2,\ldots,n+1)
\ \xrightarrow[p_1\rightarrow0]{\vspace{1mm}}
\ 0\,,
\end{equation}
where $p_1$ denotes the vanishing scalar momentum.

Moving on to double-soft emission, several different situations have to be
distinguished, which are labelled by the number of common indices between
the sets $\lbrace I_1,I_2,I_3,I_4\rbrace$ and 
$\lbrace I_5,I_6,I_7,I_8\rbrace$ in \eqn{eqn:Cartan}.
Four common indices allow
the creation of an $SU(8)$ singlet, corresponding to the emission of a
single soft graviton.  This case is not interesting because $[X,X]$
vanishes.  Similarly, if the scalars share one or two indices, the
situation corresponds to a single soft limit in one of the subamplitudes
generated by the BCFW recursion relations; thus this limit vanishes, and
does not probe the commutator in \eqn{eqn:Cartan}.  Another way to see the
vanishing is to reconsider \eqn{eqn:Cartan} explicitly: there are simply
not enough indices to saturate the right-hand side.  The only interesting
configuration occurs if the two scalars $X_1$ and $X_2$ agree on exactly
three of their indices. This result is in accordance with the commutation
relation~(\ref{eqn:Cartan}), where three equal indices are necessary for the
commutator of two noncompact generators to yield a result proportional to
an $SU(8)$ generator.

Performing an explicit calculation of an $(n+2)$-point supergravity
tree amplitude $M_{n+2}$ containing two scalars sharing three indices
and considering the double-soft limit on $X_1$ and $X_2$
results in the double-soft limit~\cite{ArkaniHamed2008gz}
\begin{equation}\label{eqn:ACK}
M_{n+2}(1,2,\ldots,n+2)
\ \xrightarrow[p_1,p_2\rightarrow0]{\vspace{1mm}}
\ \frac{1}{2}
\sum\limits_{i=3}^{n+2}\frac{p_i\cdot(p_2-p_1)}{p_i\cdot(p_1+p_2)}
T(\eta_i)M_n(3,4,\ldots,n+2)\,,
\end{equation}
where
\begin{equation}\label{eqn:su8generator}
T(\eta_i)^J_K\ =\ T\left([X^{I_1\ldots I_4},X_{I_5\ldots
I_8}]\right)_K^J
\ =\ \ve^{I_1I_2I_3I_4K}_{I_5I_6I_7I_8J}\times\eta_{iK}\pd_{\eta_{iJ}}
\end{equation}
acts on $(M_n)_J^K$; the $n$-point amplitude $M_n$ has open $SU(8)$
indices due to the particular choice of indices of the scalars. 
Again, $\ve^{I_1I_2I_3I_4K}_{I_5I_6I_7I_8J}=1,-1,0$ if the
upper index set is an even, odd or no permutation of the lower set. 

The Grassmann
variables $\eta_{iA}$ in the argument of \eqn{eqn:su8generator} refer
to the description of an amplitude in the so-called on-shell superspace
formalism~\cite{Nair1988bq}.  They are a set of $8n$ anticommuting
objects, where the index $i=1,\ldots,n$ numbers the particles
and $A$ is an $SU(8)$ index. Using these variables, one can write down
a generating functional for MHV amplitudes in
supergravity~\cite{Bianchi2008pu},
\begin{equation}\label{eqn:generatingfunctional}
\Omega_n= \frac{1}{256}
\frac{M_n(B^-_1,B^-_2,B^+_3,B^+_4,\ldots,B^+_n)}{\<12\>^8}
\prod\limits_{A=1}^{8}\sum\limits_{i,j=1}^{n}\<ij\>\eta_{iA}\eta_{jA}\,,
\end{equation}
where $B^\pm$ are positive and negative helicity gravitons.
We employ the spinor-product notation $\<ij\> = \<p_i^-|p_j^+\>$,
$[ij] = \<p_i^+|p_j^-\>$, normalized by $\<ij\>[ji]=2p_i\cdot p_j$,
where $|p_i^\pm\>$ are massless Weyl spinors.
Particle states of the $\CN=8$ multiplet
can be identified with derivatives with respect to the anticommuting
variables,
\begin{align}
  &1 \leftrightarrow B_i^+
  \qquad\frac{\pd}{\pd\eta_{iA}}\leftrightarrow F_i^{A+}
  \qquad\cdots
  \qquad\frac{\pd^4}{\pd\eta_{iA}\pd\eta_{iB}\pd\eta_{iC}\pd\eta_{iD}}
  \leftrightarrow X^{ABCD}
  \qquad\cdots\nonumber\\
  &\cdots
\quad-\frac{1}{7!}\ve_{ABCDEFGH}
\frac{\pd^7}{\pd\eta_{iB}\pd\eta_{iC}\ldots\pd\eta_{iH}}
\leftrightarrow F_{iA}^-
\quad\!\cdots\quad\!
\frac{1}{8!}\ve_{ABCDEFGH}
\frac{\pd^8}{\pd\eta_{iA}\pd\eta_{iB}\ldots\pd\eta_{iH}}
 \leftrightarrow
  B_i^-\,,\nonumber\\
&~  \label{derivclass}
\end{align}
where the number of $\eta$'s is connected to the helicity of the state, and 
$F^\pm$ denote gravitini of either helicity.  Acting with these 
operators on the generating functional~(\ref{eqn:generatingfunctional}), 
one obtains the correct expressions for the corresponding
component amplitudes, which automatically obey the MHV
supersymmetry Ward identities.
For example a two-gravitino two-graviton amplitude will read:
 \begin{align}
 \<F^{5+}\,F_5^-\,B^+\,B^-\>
&\equiv\ M_4(F_1^{5+},F_{2,5}^-,B_3^+,B_4^-)
\nonumber\\
&=-\left(\frac{\pd}{\pd\eta_{15}}\right)
 \left(\frac{1}{7!}\ve_{12345678}\frac{\pd^7}
   {\pd\eta_{21}\ldots\pd\eta_{24}\pd\eta_{26}\ldots\pd\eta_{28}}\right)
\nonumber\\
 &\quad\times\left(\frac{1}{8!}\ve_{12345678}
\frac{\pd^8}{\pd\eta_{41}\ldots\pd\eta_{48}}\right)
 \Omega_4\,.\label{eqn:su8rotation}
 \end{align}
As we will see below, the $SU(8)$ generator~\eqref{eqn:su8generator} will
act consistently on the remnant of the six-point amplitude represented in
the above formalism.

In the double-soft limit~(\ref{eqn:ACK}), the amplitude with two soft
scalars sharing three indices becomes a sum of amplitudes with only hard
momenta; in each summand one leg gets $SU(8)$ rotated by an amount
depending on its momentum.  This relation has been proven by ACK at
tree-level for pure $\CN=8$ supergravity.  Here we will construct a
suitable $\a'$-corrected amplitude, derived from an action containing the
supersymmetrized version of the $R^4$ term, and then take the double-soft
limit numerically in order to test the $E_{7(7)}$ invariance of this term.

In order to do so, we will first have a look at string theory corrections to
field theory amplitudes in the next section, before we set up the actual
calculation in section~\ref{sec:settingupcalc}.


\section{String theory corrections to field theory amplitudes}\label{sec:corr}

Tree amplitudes for Type I open and Type II closed string theory have been
computed and expanded in $\a'$ for various collections of external states.
The leading terms in the low-energy effective action
are $\CN=4$ SYM and $\CN=8$ supergravity, respectively~\cite{Green1982sw}.
Indeed, in the zero Regge slope limit $(\a'\rightarrow 0)$, 
the string amplitudes agree with the corresponding field theory results.

Expanding the string theory amplitude further in $\a'$ yields corrections
to the field-theoretical expressions, which can be summarized by a series
of local operators in the effective field theory. Terms which have to be
added to the $\CN=4$ SYM and $\CN=8$ supergravity actions in order to
reproduce the $\a'$ corrections have been identified for low orders in
$\a'$. In particular, the first nonzero string correction to the action of
$\CN=8$ supergravity is the supersymmetrized version of the possible $R^4$
counterterm~(\ref{eqn:R4}) discussed above~\cite{Gross1986iv}.

The next subsection reviews properties of amplitudes in maximally
supersymmetric field theories.   Some recent computations of 
string theory amplitudes and their low-energy expansions are discussed
in the following subsection.


\subsection{Tree-level amplitudes in $\CN=4$ SYM and $\CN=8$ Supergravity}

A general amplitude in $\CN=4$ SYM can be color-decomposed as 
\begin{equation}
 \CA_n^\SYM(1,2,\ldots,n)\ =\ g^{n-2}_{Y\!M}
\sum\limits_{\s\in S_n/\ZZ_n}\Tr(T^{a_{\s(1)}}\cdots T^{a_{\s(n)}})
\,A^\SYM_n(\s(1),\s(2),\ldots,\s(n)),
\label{treeSYMcolor}
\end{equation}
where the summation is over all $(n-1)!$ non-cyclic permutations of
$i=1,2,\ldots,n$.  The number $i$ is understood as a collective label 
for the momentum $p_i$ and helicity $h_i$ of particle $i$, {\it e.g.}
$1\equiv(p_1,h_1)$, and the $T^{a_i}$ are matrices in the fundamental
representation of the Yang-Mills gauge group $SU(N_c)$, normalized to
$\Tr(T^aT^b)=\delta^{ab}$.

The gauge-invariant subamplitudes $A^\SYM_n$ are independent of 
the color structure and can be shown to
exhibit the following properties~\cite{Mangano1990by}:
\begin{itemize}
 \item invariance under cyclic permutations:
$A_n^\SYM(1,2,\ldots,n)\,=\,A^\SYM_n(2,3,\ldots,n,1)$
 \item reflection identity:
$A_n^\SYM(1,2,\ldots,n)\,=\,(-1)^nA_n^\SYM(n,n-1,\ldots,2,1)$
 \item photon decoupling (or dual Ward) identity: 
\begin{align}\label{eqn:identities}
  &A_n^\SYM(1,2,3,\ldots,n)+A_n^\SYM(2,1,3,\ldots,n)+A_n^\SYM(2,3,1,\ldots,n)
 \nonumber\\
 &\hskip5.8cm +\,\cdots\,+A_n^\SYM(2,3,\ldots,1,n)\ =\ 0.
\end{align}
\end{itemize}
In addition, amplitudes in maximally supersymmetric theories are classified by
their helicity structure. Employing supersymmetric Ward identities (see 
section~\ref{sec:SWI}), pure-gluon amplitudes with helicity structure
$(\pm+\cdots+)$ can be shown to vanish~\cite{Grisaru1976vm,Grisaru1977px}.
The simplest nonvanishing configurations
$(--+\cdots+)$ are called maximally helicity violating (MHV) 
amplitudes.  In the case that all external legs are gluons $g^\pm$,
they are given by~\cite{Parke1986gb}: 
\begin{equation}\label{eqn:MHVpuregluon}
A_n^\SYM(g_1^-,g_2^-,g_3^+,\ldots,g_n^+)
\ =\ i \, \frac{\<12\>^4}{\<12\>\<23\>\cdots\<n1\>}\,.
\end{equation}
The simplicity of the MHV sector is also expressed in the relations between
different MHV amplitudes: any MHV amplitude is related directly 
to the pure-gluon one by supersymmetric Ward identities
(see section \ref{sec:SWI}), so that the knowledge of
\eqn{eqn:MHVpuregluon} determines the complete set of MHV amplitudes.

While in the four- and five-point case the only nonvanishing
configurations are MHV (or anti-MHV), the advent of a sixth leg introduces
a new class of helicity structures, the so-called next-to-MHV (NMHV)
amplitudes. Here it is necessary to distinguish three different helicity
orderings
\begin{equation}\label{eqn:helord}
 X:\;(---+++)\quad\qquad Y:\;(--+-++)\quad\qquad Z:\;(-+-+-+)\,.
\end{equation}
Expressions for the amplitudes are distinct for the different orderings $X$,
$Y$ and $Z$. However, there is no procedural difference in deriving 
the expressions, so we will generally illustrate the amplitudes and
supersymmetry relations for the helicity configuration $X$.
Explicit results for all six-point pure-gluon
NMHV amplitudes can be found in ref.~\cite{Mangano1990by}, for example.
More compact expressions result 
from use of the BCFW recursion relations.  Using these relations, a
prescription for determining all tree-level amplitudes in $\CN=4$ SYM
from superconformal invariants has been derived~\cite{Drummond2008cr}.

We note that the supersymmetric Ward identities, reflection symmetry and
cyclic invariance --- as well as parity, or spinor conjugation ---
relate amplitudes within a certain NMHV helicity ordering
only ($X$, $Y$ or $Z$).  On the other hand, the photon decoupling
identity is an example of a relation among amplitudes featuring
different helicity orderings. 

Next we turn to amplitudes in $\CN=8$ supergravity.  In this case, 
the color trace, which forces particles in gauge-theory subamplitudes
to remain in a certain cyclic order, does not exist.  
Instead, supergravity amplitudes are symmetric under exchange of particles 
with the same helicity.  We write the full amplitude
$\CM_n^\SUGRA(1,2,\ldots,n)$ as
\begin{equation}
 \CM_n^\SUGRA(1,2,\ldots,n)
\ =\ \left(\frac{\kappa}{2}\right)^{(n-2)}\,M_n^\SUGRA(1,2,\ldots,n),
\end{equation}
where only the gravitational coupling constant $\kappa=\sqrt{32\pi G_N}$
has been removed from $M_n^\SUGRA$.
The four- and five-point MHV amplitudes for gravitons $B^\pm$ are given by
\cite{Berends1988zp}
\begin{align}\label{eqn:GravTreeFourFive}
\begin{split}
M_4^\SUGRA(B_1^- , B_2^- , B_3^+ , B_4^+)
&=i \, \<12\>^8 \frac{ [12]}{\<34\> \, N(4)}\,,\\
M_5^\SUGRA(B_1^- , B_2^- , B_3^+ , B_4^+ , B_5^+)
&=i \, \<12\>^8 \frac{\ve(1,2,3,4)}{N(5)} \,, \cr
\end{split}
\end{align}
where
\begin{equation}\label{eqn:LeviCivitaDef}
\ve(i,j,m,n)=4i\ve_{\m\n\r\s} p_i^\m p_j^\n p_m^\r
p_n^\s=[ij]\<jm\>[mn]\<ni\>-\<ij\>[jm]\<mn\>[ni]\,
\end{equation}
and
\begin{equation}\label{eqn:NGravDef}
 N(n) \equiv \prod_{i=1}^{n-1} \prod_{j=i+1}^n \<ij\> \,.
\end{equation}
The higher-point MHV graviton amplitudes were first written down in
ref.~\cite{Berends1988zp}. Explicit expressions for other helicity
configurations are rare.   However, in ref.~\cite{Drummond2009ge} a
prescription is given how to calculate any $\CN=8$ supergravity tree-level
amplitude by employing ``gravity subamplitudes'', BCFW recursion relations,
and superconformal invariants~\cite{Drummond2008vq,Drummond2008cr}.

In the on-shell superspace formalism introduced above, the determination
of the type of amplitude away from those containing gluons (gravitons)
exclusively can be done by counting derivatives acting on the appropriate
generating functional. While $8\,(16)$ derivatives are necessary for MHV
amplitudes in $\CN=4$ SYM ($\CN=8$ supergravity), there are $12\,(24)$
derivatives in the NMHV sector.


\subsection{Amplitudes in open and closed string theory}

Open-string tree amplitudes $\CA_n$ have the same color
decomposition~(\ref{treeSYMcolor}), with $A^\SYM_n$
replaced by the color-ordered string subamplitude $A_n$.
At the four-point level, the two subamplitudes are related by the
Veneziano formula,
\begin{align}\label{eqn:s4point}
A_4(1^-,2^-,3^+,4^+)
&=\ V^{(4)}(s_1,s_2) \, A_4^\SYM(1^-,2^-,3^+,4^+) \nonumber\\
&=\ \frac{\Gamma(1+s_1)\Gamma(1+s_2)}{\Gamma(1+s_1+s_2)}
\, A_4^\SYM(1^-,2^-,3^+,4^+)\,.
\end{align}
The above expression is given in terms of kinematical invariants defined via 
\begin{equation}\label{eqn:kininv}
[\![i]\!]_n=\a'\,(p_i+p_{i+1}+\cdots+p_{i+n})^2\,,
\qquad s_j=s_{j\,j+1}=[\![j]\!]_1\,,
\qquad t_j=[\![j]\!]_2 \,,
\end{equation}
which are $s_1=[\![1]\!]_1=s_{12}=2\a' p_1\cdot p_2$ and
$s_2=[\![2]\!]_1=s_{23}=2\a' p_2\cdot p_3$ on-shell. 
Expanding the form-factor $V^{(4)}$ in powers of $\a'$ one finds
\begin{equation}
 V^{(4)}(s_1,s_2)\ =\ 1-\zeta(2) s_1 s_2+\zeta(3) s_1 s_2
(s_1+s_2)+\CO(\a'^4),
\end{equation}
where the leading correction to the pure Yang-Mills amplitude arises
from the interaction term of four gauge field-strength
tensors~\cite{Green1981xx,Schwarz1982jn,Tseytlin1986ti}.

The full open string amplitude is quite simple in the four-point
case~\eqref{eqn:s4point}.  On the other hand, its generalizations to 
more external legs turn out to involve generalized
hypergeometric functions~\cite{Oprisa2005wu}. Any $n$-point open string
amplitude can be expressed in terms of $(n-3)!$ hypergeometric basis
integrals.  Expanding those functions in powers of $\a'$ yields
expressions for the string-corrected five- and six-point MHV amplitudes,
\begin{align}
A_5&=\left[V^{(5)}(s_j) - \frac{i\,\a'^2}{2}
\,\ve(1,2,3,4)P^{(5)}(s_j)\right]A_5^\SYM \,, \nonumber\\
A_6&=\left[V_6^\open(s_j,t_j)-\frac{i\,\a'^2}{2}
\,\sum\limits_{k=1}^{5}\ve_k P_k^{(6)}(s_j,t_j)\right]A_6^\SYM\,,
\label{56MHVopen}
\end{align}
where
\begin{equation}
\ve_1=\ve(2,3,4,5),
\quad\ve_2=\ve(1,3,4,5),
\quad\ve_3=\ve(1,2,4,5),
\quad\ve_4=\ve(1,2,3,5),
\quad\ve_5=\ve(1,2,3,4)\,.
\end{equation}
Expansions in $\a'$ are given by~\cite{Stieberger2006te}
\begin{align}
V^{(5)}(s_i)\,=\ &
1-\frac{\zeta(2)}{2}(s_1s_2+s_2s_3+s_3s_4+s_4s_5+s_5s_1)\nonumber\\
&+\frac{\zeta(3)}{2}
\big(s_1^2s_2+s_2^2s_3+s_3^2s_4+s_4^2s_5+s_5^2s_1
+s_1s_2^2+s_2s_3^2+s_3s_4^2+s_4s_5^2+s_5s_1^2\nonumber\\
&\qquad\qquad+s_1s_3s_5+s_2s_4s_1+s_3s_5s_2+s_4s_1s_3+s_5s_2s_4\big)
+\CO(\a'^4)\,,\\
P^{(5)}(s_i)\,=\ & \zeta(2)-\zeta(3)(s_1+s_2+s_3+s_4+s_5)+\CO(\a'^2)\,,
\end{align}
and explicit expressions for $V^{(6)}$ and $P_k^{(6)}$ can be found
in the same reference.

Stieberger and Taylor have pushed the calculations even
further~\cite{Stieberger2007am}.  In the process of determining all 
pure-gluon NMHV six-point amplitudes, they computed the following
additional auxiliary amplitudes for the helicity configuration $X$
defined in \eqn{eqn:helord}:
\begin{equation}
\<\phi^-\phi^-\phi^-\phi^+\phi^+\phi^+\>\,,\quad
\<\phi^-\phi^-\la^-\la^+\phi^+\phi^+\>\,,\quad\text{and}
\quad\<\phi^-\phi^-g^-g^+\phi^+\phi^+\> \,,
\label{STamps}
\end{equation}
as well the analogous quantities for $Y$ and $Z$.
Here $\la$ denotes a gluino and $\phi$ a scalar.
In order to get an impression of the complexity of the result, we provide
the pure-gluon NMHV six-point amplitude in helicity configuration
$X$~\cite{Stieberger2007am}, which will be expressed employing the 
following kinematic variables:
\begin{equation}
\a_X = - \, [12]\<34\>[\,6|X|5\>~,
\qquad \b_X = [12]\<45\>[\,6|X|3\>~,
\qquad \g_X = [61] \<34\>[2|X|5\>\,,
\end{equation}
where $X \equiv p_6+p_1+p_2$.  The subamplitude reads\footnote{%
Note the shifted ordering of helicities compared
to \eqn{eqn:helord}. A cyclic shift $(1,2,3,4,5,6)\rightarrow(3,4,5,6,1,2)$
has to be performed in order to match the results analytically with
ref.~\cite{Stieberger2007am}.}
\begin{align}\label{eqn:AmpX}
&A_6(g^+_1,g^+_2,g^-_3,g^-_4,g^-_5,g^+_6)=\nonumber\\
&\hskip1cm
\frac{\a'^5}{s_5}\left(N^X_1\frac{\a_X^2}{s_1^2s_3^2}
+ N^X_2\frac{\b_X^2}{s_1^2} + N^X_3\frac{\g_X^2}{s_3^2}
+ N^X_4\frac{\a_X\b_X}{s_1^2s_3} + N^X_5\frac{\a_X\g_X}{s_1s_3^2}
+ N^X_6\frac{\b_X\g_X}{s_1s_3}\right)\ ,
\end{align}
where the expansion of the functions $N^X$ to $\CO(\a'^2)$ is:
\begin{align}
N^X_1&=-\zeta(2)\ s_1s_3+\ldots\ ,\nonumber\\
N^X_2&=\frac{s_1}{s_2s_4t_1}-\zeta(2)\
\left(\frac{s_1s_6}{s_2s_4}+\frac{s_1^2}{s_4t_1}+\frac{s_1s_5}{s_2t_1}
\right)+\ldots\ ,\nonumber\\
N^X_3&=\frac{s_3}{s_2s_6t_2}-\zeta(2)\
\left(\frac{s_3s_4}{s_2s_6}+\frac{s_3s_5}{s_2t_2}+\frac{s_3^2}{s_6t_2}
\right)+\ldots\ ,\nonumber\\
N^X_4&=\zeta(2)\ \left(\frac{s_1t_2}{s_2}+\frac{s_1t_3}{s_4}\ \right)
+\ldots\ ,\nonumber\\
N^X_5&=\zeta(2)\ \left(\frac{s_3t_1}{s_2}+\frac{s_3t_3}{s_6}\ \right)
+\ldots\ ,\nonumber\\
N^X_6&=\frac{t_3}{s_2s_4s_6}+\zeta(2)\
\left(\frac{s_1+s_3-s_5}{s_2}-\frac{t_1t_3}{s_2s_4}-
\frac{t_2t_3}{s_2s_6}-\frac{t_3^2}{s_4s_6}\right)+\ldots\;.
\end{align}

The low-energy limit of closed Type II string theory in four dimensions
is $\CN=8$ supergravity.  The first correction to the low-energy effective
action can be determined from the expression for the closed string
four-point amplitude, or Virasoro-Shapiro amplitude,
\begin{align}
M_4(1^-,2^-,3^+,4^+) &= V^{(4)}_\closed(s_1,s_2) \, 
M_4^\SUGRA(1^-,2^-,3^+,4^+)\nonumber\\
&= \frac{\Gamma(1+s_1)\Gamma(1+s_2)\Gamma(1-s_1-s_2)}
{\Gamma(1-s_1)\Gamma(1-s_2)\Gamma(1+s_1+s_2)}
\, M_4^\SUGRA(1^-,2^-,3^+,4^+)\,.
\label{VirasoroShapiro}
\end{align}
The expansion of $V^{(4)}_\closed$ has the first nonvanishing
correction at $\CO(\a'^3)$,
\begin{equation}\label{eqn:closedstringexpansion}
 V^{(4)}_\closed(s_1,s_2)
\ =\  1 + 2 \,\zeta(3)\, s_1s_2(s_1+s_2)+\CO(\a'^4)\,,
\end{equation}
which corresponds to a supersymmetrized version of \eqn{eqn:R4} in the low
energy effective action~\cite{Gross1986iv}. In other words, keeping terms
up to order $\CO(\a'^3)$ in the closed-string amplitudes is equivalent
to working with a theory whose effective action is given by
\begin{equation}\label{eqn:corraction}
 S_{\text{corr}}=\int\,\diff^4x\sqrt{-g}(\CR+\a'^3\CR^4)+\CO(\a'^4)\,.
\end{equation}

While $\a'$-corrected six-point amplitudes in open string theory ($\CN=4$
SYM) are already very cumbersome to calculate, the situation is even worse
for closed string theory ($\CN=8$ supergravity).  For higher-point tree
amplitudes it is therefore more convenient to rely on the KLT
relations, which express closed string amplitudes as simple quadratic
combinations of open string amplitudes.

Several different cyclic orderings of the open string amplitudes are
required as input to the KLT relations.  Fortunately, there are several 
open string amplitudes available.  In particular,
a couple of six-point NMHV amplitudes have been
computed~\cite{Stieberger2007am}\footnote{We are grateful to Stephan 
Stieberger and Tomasz Taylor for providing us with expressions for 
the amplitudes from ref.~\cite{Stieberger2007am} through order $\a'^3$.}, 
which will serve below as input to the calculation of a suitable
$\a'$-corrected $\CN=8$ supergravity amplitude. 


\section{Setting up the calculation}\label{sec:settingupcalc}

Arkani Hamed, Cachazo and Kaplan have proven~\eqn{eqn:ACK}
analytically, by employing BCFW recursion relations for $\CN=8$ supergravity
with $E_{7(7)}$ realized on-shell. Because invariance under $E_{7(7)}$ is a
necessary condition for the relation to be valid, \eqn{eqn:ACK} provides a
useful tool for testing other theories, or operators,
for their symmetry properties under $E_{7(7)}$.
In particular, if the double-soft limit of all $(n+2)$-point amplitudes
derived from \eqn{eqn:corraction} coincides with the $SU(8)$ rotated sum
of the corresponding $n$-point amplitudes, that would be strong evidence
that $E_{7(7)}$ symmetry does not restrict the appearance of $\CR^4$ as a
counterterm in $\CN=8$ supergravity.

The analytical approach that ACK used to prove \eqn{eqn:ACK} does not hold
for the $\a'$-corrected $\CN=8$ amplitudes.  Higher-dimension operators
lead to poorer large-momentum behavior, so that amplitudes shifted by
large complex momenta will not fall off fast enough for the BCFW recursion
relations to be valid.  Instead we will find explicit (if lengthy)
expressions for suitable and available string theory amplitudes, 
from which the $\a'$-corrected amplitudes corresponding to
\eqn{eqn:corraction} can be deduced, and their double-soft limits
inspected (numerically).

After we give a short introduction to the KLT relations in
subsection~\ref{sec:KLT}, we will explore the
constraints on the $\a'$-corrected $\CN=8$ supergravity amplitude
originating from the double-soft limit relation~(\ref{eqn:ACK})
in subsection \ref{sec:suitamp}.  Appropriate $\CN=8$ amplitudes
will be identified and decomposed into $\CN=4$ SYM matrix elements using
the KLT relations.  The required ($\a'$-corrected) $\CN=4$ SYM
matrix elements can be related to the available open string amplitudes by
carefully examining the NMHV supersymmetric Ward identities.
In subsections \ref{sec:SWI} and \ref{sec:N1SWI}, the $\CN=1$ supersymmetric
Ward identities will be reviewed in detail and used to finally obtain
expressions for the $\CN=4$ amplitudes, which serve as input to the KLT
relations, in section \ref{sec:calculation}.


\subsection{KLT relations}\label{sec:KLT}

Tree-level amplitudes in closed and open string theories are linked by
the KLT relations~\cite{KLT}, which arise from the fact that any
closed-string vertex operator can be represented as a product of two
open-string vertex operators,
\begin{equation}\label{eqn:stringvertexdecomp}
V^{\rm closed}(z_i,\bar{z}_i) = V_{\rm left}^{\rm open}(z_i)\,
\overline{V}_{\rm right}^{\rm  open}(\bar{z}_i) \,.
\end{equation}
While in the closed-string amplitude the insertion points
$z_i,\,\bar{z}_i$ of vertex operators are integrated over a
two-sphere, in the open-string case the real $z_i$ are integrated over
the boundary of a disk. Thus the closed-string integrand equals the
product of two open-string integrands. KLT related the two sets of
string amplitudes by evaluating the closed-string integrals via a
contour deformation in terms of the open-string integrals.

The KLT relations for four-, five- and six-point amplitudes are 
\begin{align}
M_4 (1,2,3,4)&= \frac{- i}{\a'\pi}
 \sin(\pi s_{12})\,A_4(1,2,3,4)\,A_4(1,2,4,3)\,,
\label{KLT4}\\
M_5(1,2,3,4,5)&=\frac{ i}{\a'^2\pi^2}\sin(\pi s_{12})\sin(\pi
s_{34})\,A_5(1,2,3,4,5)\,A_5(2,1,4,3,5)\nonumber\\
&\qquad\qquad +\CP(2,3)\,,\label{KLT5}\\
M_6(1,2,3,4,5,6)&=\frac{- i}{\a'^3\pi^3}
\sin(\pi s_{12})\sin(\pi s_{45})\,A_6(1,2,3,4,5,6)\nonumber\\
&\qquad \qquad
\times \left[\sin(\pi s_{35})\,A_6(2,1,5,3,4,6)+\sin(\pi(s_{34}+
s_{35}))\,A_6(2,1,5,4,3,6)\right]\nonumber\\
&\qquad\qquad +\CP(2,3,4) \,,
\label{KLT6}
\end{align}
where ``$+\CP$'' indicates a sum over the $m!$ permutations of the 
$m$ arguments of $\CP$.
Formulae for higher-point amplitudes can be derived
straightforwardly~\cite{KLT}.  In the field-theory ($\a'\to0$) limit,
a closed form has been obtained for all $n$~\cite{Bern1998sv}.

The above equalities are exact relations between string theory amplitudes,
and so they are valid order by order in $\a'$.  In order to calculate the
string correction to an $\CN=8$ supergravity amplitude at a certain order
in $\a'$ from known $\a'$-corrected expressions in $\CN=4$ SYM, one has to
determine all combinations of terms from the expansions of the amplitudes
and the sine functions, whose multiplication results in the correct power
of $\a'$. For instance, the second-order correction to the 
five-point amplitude in supergravity
corresponds to terms of $\CO(\a'^4)$, due to the
prefactor of $\sfrac{1}{\a'^2}$.  Taking the absence of first-order
corrections to $\CN=4$ SYM amplitudes into account, four combinations
have to be considered in \eqn{KLT5}, according to
table~\ref{table:KLTcomb},
\begin{table}
\begin{center}
 \begin{tabular}{|c|c|c|c|}
  \hline
  $\sin(\pi s_{12})$&$\sin(\pi s_{34})$&$A_5(1,2,3,4,5)$&$A_5(2,1,4,3,5)$\\
  \hline\hline
  $\CO(\a'^1)$&$\CO(\a'^1)$&$\CO(\a'^0)$&$\CO(\a'^2)$\\
  $\CO(\a'^1)$&$\CO(\a'^1)$&$\CO(\a'^2)$&$\CO(\a'^0)$\\
  $\CO(\a'^3)$&$\CO(\a'^1)$&$\CO(\a'^0)$&$\CO(\a'^0)$\\
  $\CO(\a'^1)$&$\CO(\a'^3)$&$\CO(\a'^0)$&$\CO(\a'^0)$\\\hline
 \end{tabular} 
\end{center}
\caption{Enumeration of the orders in $\a'$ required from various
factors, in order to compute the five-point closed-string amplitude to
$\CO(\a'^2)$ using the KLT relations.
\label{table:KLTcomb}}
\end{table}
yielding
\begin{align}
 M_5^{\CO(\a'^2)}=\frac{i s_{12}s_{34}}{\a'^2}
\left[ A_5^\SYM(1,2,3,4,5)\,A_5^{\CO(\a'^2)}(2,1,4,3,5)
     + A_5^{\CO(\a'^2)}(1,2,3,4,5)\,A_5^\SYM(2,1,4,3,5)
\right.
\nonumber\\
\left.\hskip-1cm - \sfrac{\pi^2}{6}(s_{12}^2+s_{34}^2)
A_5^\SYM(1,2,3,4,5)\,A_5^\SYM(2,1,4,3,5)\right
]+\CP(2,3)\,.
\end{align}
The above expression can be shown to vanish analytically, in accordance
with the higher-point generalization of \eqn{eqn:closedstringexpansion},
or alternatively \eqn{eqn:corraction}, the statement that the first
correction to the closed-string effective action is at $\CO(\a'^3)$.

Although the KLT relations are often applied to pure-graviton and pure-gluon
amplitudes, their use is not limited to these scenarios. Any pair of
consistent open-string amplitudes is related to an amplitude in closed
string theory and vice versa. Considering the combination of two
open-string vertex operators into a closed one in
\eqn{eqn:stringvertexdecomp}, one can immediately determine which type of
particle has to appear at a certain position on the supergravity side by
adding the helicities and combining the indices, according to the
tensor-product decomposition of the Fock space,
\begin{equation}
 [\CN=8]\ \leftrightarrow\ [\CN=4]_L\otimes[\CN=4]_R \,.
\end{equation}
Somewhat remarkably, the opposite statement is true as well:
given a certain operator, corresponding to a particular state in
$\CN=8$ supergravity, the helicity, global symmetry properties,
and the consistent action of supercharges in either of the
theories are sufficient to unambiguously determine the decomposition into
$\CN=4$ SYM states~\cite{Bianchi2008pu}. The decompositions relevant
for the calculation to follow are
\begin{align}\label{eqn:KLTdecomposition}
B^+\,&=\,g^+\tilde{g}^+\,,\quad&F^{a+}\,&=\,\la^{a+}\tilde{g}^+\,,\quad&
F^{r+}\,&=\,g^+\tilde{\la}^{r+}\,,\quad\nonumber\\
B^-\,&=\,g^-\tilde{g}^-\,,\quad&F_a^-\,&=\,\la_a^-\tilde{g}^-\,,\quad
&F^-_r\,&=\,g^-\tilde{\la}_r^-\,,\quad\nonumber\\
X^{abcd}\,&=\,\ve^{abcd}\,g^-\,\tilde{g}^+\,,\quad
&X^{abcr}\,&=\,\ve^{abcd}\,\la_d^-\,\tilde{\la}^{r+}\,,\quad
&X^{abrs}\,&=\,\phi^{ab}\,\tilde{\phi}^{rs}\,,
\nonumber\\
X_{abcd}\,&=\,\ve_{abcd}\,g^+\,\tilde{g}^-\,,\quad
&X_{abcr}\,&=\,\ve_{abcd}\,\la^{d+}\,\tilde{\la}_r^-\,,\quad
&X_{abrs}\,&=\,\phi_{ab}\,\tilde{\phi}_{rs}\,,
\end{align}
where capital letters $B,\,F,\,X$ denote the graviton, gravitino and
scalar particle in $\CN=8$ supergravity and $g,\,\la,\, \phi$ the gluon,
gluino and scalar in $\CN=4$ SYM. Quantities with indices $a,b,\ldots$
correspond to the first $SU(4)$, while quantities with a tilde and indices
$r,s,\ldots$ are in the second $SU(4)$.  (In particular, $\tilde{g}$ does
{\it not} denote a gluino!)  Finally, the superscripts $+$ and
$-$ mark the helicity signature.


\subsection{Choosing a suitable amplitude}\label{sec:suitamp}

The simplest scenario one might think of, in order to test the double-soft
scalar limit relation~\eqref{eqn:ACK}, would be to start with a five-point
amplitude, which in turn would lead to a sum of three-point
amplitudes on the right-hand side of the relation. Three-point amplitudes
are special as they require a setup with complex momenta in order to be
non-trivial. However, here we have to take another constraint into
account: we want to test amplitudes that receive nonvanishing 
corrections from the $\CR^4$ term.  Because the interactions originating
in this counterterm candidate start at the four-point level,
it is not sufficient to consider three-point amplitudes.

Therefore we will have to consider a six-point amplitude,
which should reduce to a sum of four-point amplitudes in the
double-soft limit.  We again require that the four-point amplitudes
on the right-hand side of \eqn{eqn:ACK} are nonvanishing, which
implies that they are MHV (or equivalently anti-MHV).
Fortunately, corrections to all MHV-amplitudes with four legs are
known up to $\CO(\a'^3)$, indeed to arbitrary orders in $\a'$, using
\eqn{VirasoroShapiro} and the MHV supersymmetry Ward identities.

On the left-hand side of \eqn{eqn:ACK} the situation is more intricate.
The four particles that appear already on the right-hand side are now
accompanied by two additional scalars.  According to \eqn{derivclass}, 
the number of $\eta$ derivatives acting on the generating functional
is increased by eight, four for each scalar, so that the 
resulting amplitude resides in the NMHV sector. In addition, the two
scalars have to share three $SU(8)$ indices, as elaborated on in
section~\ref{sec:coset}. Sorting out the distribution of the scalars'
indices into two $SU(4)$ subgroups, there are finally five possible
distinct choices\footnote{Another five combinations can be
obtained by switching the left and right $SU(4)$.}
satisfying the constraints.  They are listed here, together with
their respective KLT decompositions
according to equation \eqref{eqn:KLTdecomposition}:
\begin{align}
\label{eqn:choice1}\<X^{abrs}X_{abrt}\cdot\;\cdot\;\cdot\;\cdot\,
\>&\ \rightarrow\ \<\phi^{ab}\,\phi_{ab}\,\cdot\;\cdot\;\cdot\;\cdot\,{\>}
_L\times\<\phi^{rs}\,\phi_{rt}\,\cdot\;\cdot\;\cdot\;\cdot\,{\>}_R,\\
\label{eqn:choice2}\<X^{abrc}X_{abrs}\cdot\;\cdot\;\cdot\;\cdot\,
\>&\ \rightarrow\ \<\ve^{abcd}\la_d^{-}\,
\phi_{ab}\,\cdot\;\cdot\;\cdot\;\cdot\,{\>}
_L\times\<\la^{r+}\,\phi_{rs}\,\cdot\;\cdot\;\cdot\;\cdot\,{\>}_R,\\
\label{eqn:choice3}\<X^{abrc}X_{abrd}\cdot\;\cdot\;\cdot\;\cdot\,
\>&\ \rightarrow\ \<\la_d^{-}\,\la^{c+}\,\cdot\;\cdot\;\cdot\;\cdot\,{\>}
_L\times\<\la^{r+}\,\la_r^{-}\,\cdot\;\cdot\;\cdot\;\cdot\,{\>}_R,\\ 
\label{eqn:choice4}\<X^{abcr}X_{abcs}\cdot\;\cdot\;\cdot\;\cdot\,
\>&\ \rightarrow\ \<\la_d^{-}\,\la^{d+}\,\cdot\;\cdot\;\cdot\;\cdot\,{\>}
_L\times\<\la^{r+}\,\la_s^{-}\,\cdot\;\cdot\;\cdot\;\cdot\,{\>}_R,\\
\label{eqn:choice5}\<X^{abcd}X_{abcr}\cdot\;\cdot\;\cdot\;\cdot\,
\>&\ \rightarrow\ \<g^-\,\la^{d+}\,\cdot\;\cdot\;\cdot\;\cdot\,{\>}_L
\times\<g^+\,\la_r^{-}\,\cdot\;\cdot\;\cdot\;\cdot\,{\>}_R\,.
\end{align}
Here the ellipses are understood to be filled with four particles such
that the $L$- and $R$-amplitudes on the right-hand side of the KLT
relation each transform as an $SU(4)$ singlet. In each of
equations~\eqref{eqn:choice3} to~\eqref{eqn:choice5} we have left out a
factor of $\ve^{abcd}\ve_{abcd}$.  Because these indices are not 
summed over, this factor is equal to unity.  Note that 
$\<X^{abcd}X_{abce}\cdot\;\cdot\;\cdot\;\cdot\,\>$ is absent because
the five $SU(4)$ indices $a,b,c,d,e$ cannot be made all distinct.

In order to proceed, we need to use supersymmetric Ward identities
to relate one of the five
decompositions~\eqref{eqn:choice1}--\eqref{eqn:choice5}
to the available open-string six-point results (see \eqn{STamps} in 
section~\ref{sec:corr}):
\begin{align}\label{eqn:availamps}
&\<g^-\,g^-\,g^-\,g^+\,g^+\,g^+\>\,,
\quad \<\phi^-\,\phi^-\,\phi^-\,\phi^+\,\phi^+\,\phi^+\>,\nonumber\\
&\<\phi^-\,\phi^-\,\la^-\,\la^+\,\phi^+\,\phi^+\>\quad
\text{and}\quad\<\phi^-\,\phi^-\,g^-\,g^+\,\phi^+\,\phi^+\>\,.
\end{align}
Supersymmetric Ward identities can be classified by
the amount of supersymmetry employed ({\it e.g.}, $\CN=1,2,4$),
as well as the number of legs and the sector (MHV, NMHV, {\it etc.})
characterizing the amplitudes. 
We deal with six-point NMHV amplitudes exclusively here.
The notation $\CN=4$ SWI will refer to the set of supersymmetric Ward 
identities relating six-point NMHV amplitudes built
from the full $\CN=4$ multiplet $(g^\pm,\la_m^\pm,\phi_n^\pm)$, where
$m=1,2,3,4$ and $n=1,2,3$.  (Note that a superscript $\pm$ on $\phi$
implies a complex field with a different index labelling from the 
real $\phi_{ab}$ used above.)
In the original article~\cite{Stieberger2007am}, $\CN=2$ supersymmetric
Ward identities served to relate the latter three amplitudes 
in \eqn{eqn:availamps} to the pure-gluon one.  So the obvious idea would
be to search in the decompositions~\eqref{eqn:choice1}--\eqref{eqn:choice5}
for one in which the amplitudes contain particles from a single $\CN=2$
multiplet (plus its CPT conjugate), 
$(g^\pm,\la_m^\pm,\phi_1^\pm)$ with $m=1,2$.

However, the third amplitude in~\eqn{eqn:availamps} contains only one
type of fermion, which points into the direction of a $\CN=1$
multiplet. Setting up the calculation employing $\CN=1$ SWI
exclusively is a bit simpler than using $\CN=2$ SWI: For six-point
NMHV amplitudes an explicit and simple solution to the $\CN=1$ SWI is
known~\cite{Bianchi2008pu, Grisaru1977px}.  (We note that very
recently the supersymmetric Ward identities in maximally
supersymmetric ${\cal N}=4$ super-Yang-Mills theory and ${\cal N}=8$
supergravity were solved, quite remarkably, for arbitrary $n$-point
N$^k$MHV amplitudes~\cite{Elvang2009wd} in terms of basis amplitudes,
in a manifestly supersymmetric form.  These results may prove very
useful in extending the considerations of this paper to greater
numbers of legs.)

Now the decompositions~\eqref{eqn:choice1} to~\eqref{eqn:choice5}
are not all equally suited to the use of an $\CN=1$ SWI.
For example, the left $SU(4)$ amplitude of \eqn{eqn:choice2} 
contains three distinct $SU(4)$ indices, $a,b,d$, thus requiring
a full $\CN=4$ multiplet.  The other four decompositions contain
amplitudes which can be constructed from SWI with less supersymmetry.
Indeed, the decomposition~\eqref{eqn:choice5} contains only one index
for the left $SU(4)$ amplitude, and one for the right one; 
this decomposition is the one we will use in this paper.
As will be explained below, it is possible to obtain everything we
need for testing the double-soft limit through \eqn{eqn:choice5}, 
by using a two-step procedure employing two different sets of $\CN=1$
SWI based on the multiplets $(g^\pm,\la^\pm)$ and $(\phi^\pm,\la^\pm)$. 

The next three subsections introduce the SWI in general, elaborate
on the $\CN=1$ SWI for $(g^\pm,\la^\pm)$ in particular, 
and then describe the analogous set of $\CN=1$ SWI for the
multiplet $(\phi^\pm,\la^\pm)$.  Then, in section~\ref{sec:calculation},
we will assemble these ingredients in order to test the $E_{7(7)}$ symmetry.


\subsection{Supersymmetric Ward identities}\label{sec:SWI}

Supersymmetric Ward identities can be derived using the fact that
supercharges annihilate the vacuum of the theory, $Q|0\>=0$, so that
\begin{equation}\label{eqn:SWIgeneral}
0 = \<\left[Q,\b_1\b_2\cdots\b_n\right]\>
 = \sum\limits_{i=1}^n\<\b_1\b_2\cdots\left
   [Q,\b_i\right]\cdots\b_n\> \,.
\end{equation}
Here the $\b_i$ are arbitrary states from the multiplet under consideration,
$Q=Q(\eta)=\<Q\eta\>$ is a supersymmetry operator, which has been
bosonized by contraction with the Grassmann variable $\eta$, and
$\<\b_1\b_2\cdots\b_n\>$ will be called the \textit{source term} for the
SWI. Source terms need to have an odd number of fermions, because
amplitudes derived by acting on terms with an even number of fermions
will vanish trivially. An immediate and standard result implied by
\eqn{eqn:SWIgeneral} is the disappearance of all amplitudes with
helicity structure $\<+++\cdots+\>$ and $\<+--\cdots-\>$. With only
little more effort one can show that maximally helicity violating
amplitudes (MHV) are related pairwise by SWI, which in turn means that
knowing one amplitude determines the whole MHV sector for a particular
number of legs~\cite{Dixon1996wi}. In the NMHV sector this is no longer
true; here each supersymmetric Ward identity relates three amplitudes,
which requires two known amplitudes in order to determine a third one. 

Stieberger and Taylor have explicitly proven for open string theory
on the disk that the forms of the supersymmetric Ward identities to all
orders in $\a'$ are identical to those in the corresponding
four-dimensional field-theoretical limit~\cite{Stieberger2007jv}. So the
exploration in the next two subsections will be valid as well for the
$\a'$-corrected amplitudes under investigation.


\subsection{$\CN=1$ supersymmetric Ward identities}
\label{sec:N1SWI}

As an example, let us investigate the set of amplitudes involving
gluons $(g^+,g^-)$ and a single pair of gluinos $(\la^+,\la^-)$
(from which we drop the $SU(4)$ index for simplicity). 
The states are related by $\CN=1$ supersymmetry via
\begin{align}
 \left[Q(\eta),g^+(p)\right]&=\left[p\eta\right]\la^+(p),\nonumber\\
 \left[Q(\eta),\la^+(p)\right]&=-\<p\eta\>g^+(p),\nonumber\\
 \left[Q(\eta),g^-(p)\right]&=\<p\eta\>\la^-(p),\nonumber\\
 \left[Q(\eta),\la^-(p)\right]&=-\left[p\eta\right]g^-(p),
\end{align}
where $Q(\eta)=\<Q\eta\>.$

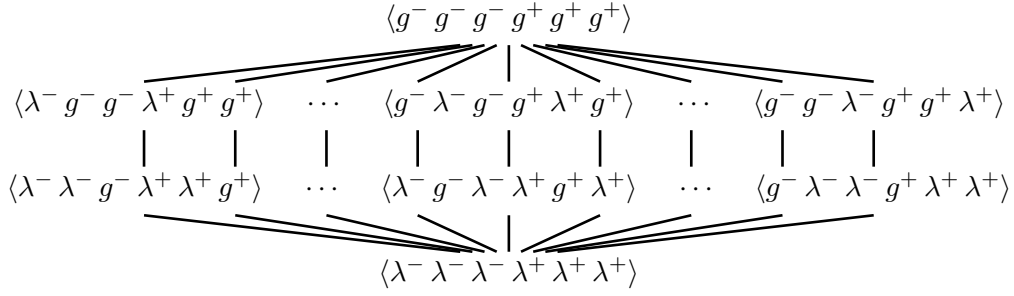
\begin{figure}[h]
\begin{center}
\psset{xunit=2.3pt,yunit=2.3pt,linewidth=1pt}
\begin{pspicture}(160,46)
 \rput[b](80,42){$\<g^-\,g^-\,g^-\,g^+\,g^+\,g^+\>$}
 \psline(80,40)(80,34)\psline(78,40)(65,34)\psline(76,40)(50,34)
\psline(74,40)(35,34)\psline(72,40)(20,34)\psline(82,40)(95,34)
\psline(84,40)(110,34)\psline(86,40)(125,34)\psline(88,40)(140,34)
 \rput[b](80,28){$\<\la^-\,g^-\,g^-\,\la^+\,g^+\,g^+\>\;
\quad\cdots\quad\;\<g^-\,\la^-\,g^-\,g^+\,\la^+\,g^+\>\;
\quad\cdots\quad\;\<g^-\,g^-\,\la^-\,g^+\,g^+\,\la^+\>$}
 \psline(80,26)(80,20)\psline(65,26)(65,20)\psline(50,26)(50,20)
\psline(35,26)(35,20)\psline(20,26)(20,20)\psline(95,26)(95,20)
\psline(110,26)(110,20)\psline(125,26)(125,20)\psline(140,26)(140,20)
 \rput[b](80,14){$\<\la^-\,\la^-\,g^-\,\la^+\,\la^+\,
g^+\>\;\quad\cdots\quad\;\<\la^-\,g^-\,\la^-\,\la^+\,g^+\,
\la^+\>\;\quad\cdots\quad\;\<g^-\,\la^-\,\la^-\,g^+\,\la^+\,\la^+\>$}
 \psline(80,6)(80,12)\psline(78,6)(65,12)\psline(76,6)(50,12)
\psline(74,6)(35,12)\psline(72,6)(20,12)\psline(82,6)(95,12)
\psline(84,6)(110,12)\psline(86,6)(125,12)\psline(88,6)(140,12)
 \rput[b](80,0){$\<\la^-\,\la^-\,\la^-\,\la^+\,\la^+\,\la^+\>$}
\end{pspicture}
\caption{Amplitudes related by $\CN=1$ supersymmetric Ward identities.}
\label{fig:N1SWI}
\end{center}
\end{figure}

For each NMHV helicity sector, there are $20$ distinct amplitudes
related by $\CN=1$ SWI: a pure-gluon amplitude, a pure-gluino amplitude,
nine two-gluino four-gluon amplitudes, and nine four-gluino two-gluon
amplitudes, as shown in figure~\ref{fig:N1SWI}.  In the following,
we assume that amplitudes are drawn from helicity configuration $X$
in \eqn{eqn:helord}.  For the two other configurations $Y$ and $Z$,
the relations are completely analogous.

Amplitudes in adjacent rows of figure~\ref{fig:N1SWI}
are related by the $\CN=1$ SWI.  Acting for example with the 
supersymmetry operator $Q(\eta)$ on the
source term $\<g^-\,g^-\,g^-\,\la^+\,g^+\,g^+\>$ yields
\begin{align}\label{eqn:pgSWI}
 \<4\eta\>\<g^-\,g^-\,g^-\,g^+\,g^+\,g^+\>
-\<1\eta\>\<\la^-\,g^-\,g^-\,\la^+\,g^+\,g^+\>&\nonumber\\
\null
-\<2\eta\>\<g^-\,\la^-\,g^-\,\la^+\,g^+\,g^+\>
-\<3\eta\>\<g^-\,g^-\,\la^-\,\la^+\,g^+\,g^+\>&=0\,,
\end{align}
which relates the pure-gluon amplitude to the two-gluino four-gluon
ones from the second row in figure~\ref{fig:N1SWI}. Due to the freedom
in choosing the two-component supersymmetry parameter $\eta$, the
result is a system of equations which has rank $2$. In order to find
all relations between the pure gluon amplitude (first row) and the
amplitudes in the second row, the action of $Q(\eta)$ on all possible
source terms featuring one gluino and five gluons,
\begin{align}
 \<\la^-\,g^-\,g^-\,g^+\,g^+\,g^+\>,\quad
 \<g^-\,\la^-\,g^-\,g^+\,g^+\,g^+\>,\quad
 \<g^-\,g^-\,\la^-\,g^+\,g^+\,g^+\>,\,\nonumber\\
 \<g^-\,g^-\,g^-\,\la^+\,g^+\,g^+\>,\quad
 \<g^-\,g^-\,g^-\,g^+\,\la^+\,g^+\>,\quad
 \<g^-\,g^-\,g^-\,g^+\,g^+\,\la^+\>,
\end{align}
has to be considered. The resulting system, linking ten amplitudes
from the first and second rows, turns out to have rank eight, thus
requiring two known amplitudes in order to derive all the others. 

Repeating the analysis for the second and third rows, there are
notably more identities to consider. They are generated by acting with
$Q(\eta)$ on any of the 18 different source terms built from three
gluinos and the same number of gluons, {\it e.g.}
$\<\la^-\,\la^-\,g^-\,g^+\,\la^+\,g^+\>$. Interestingly this system
connecting $18$ unknown amplitudes is of rank $16$, meaning that again
two amplitudes have to be known in order to fix all the others.

Finally, the relations between the third row and the pure-gluino
amplitude (fourth row) mirror the situation found for the top of the
diagram and are also of rank eight.

Combining all of the above into one large system of equations, the
total rank of the supersymmetric Ward identities pictured in
figure~\ref{fig:N1SWI} turns out to be $18$. So, given any two of the
$20$ distinct amplitudes, one can calculate any other from this set
employing the complete collection of $\CN=1$ SWI. The corresponding
result has already been found by Grisaru and Pendleton in the context
of $\CN=1$ supergravity~\cite{Grisaru1977px}, and recast recently in
modern spinor-helicity form~\cite{Bianchi2008pu}.

More explicitly, any two-gluino four-gluon amplitude $F_{i,I}$, with
the gluinos situated at positions $i$ and $I$, can be expressed in
terms of the pure-gluon and pure-gluino amplitudes as
\begin{equation}\label{eqn:BEFsol}
F_{i,I} =
\frac{4\<Ij\>[ij]\<g^-g^-g^-g^+g^+g^+\>-\ve_{ijk}\<jk\>\ve_{IJK}[JK]
\<\la^-\la^-\la^-\la^+\la^+\la^+\>}
{-2\sum_{m,n\in\lbrace i,j,k \rbrace}\<mn\>[nm]}\,,
\end{equation}
where $i,j,k$ and $I,J,K$ mark the set of negative and positive helicity
particles respectively, and the numerator contains implicit sums over
$j,k,J,K$.  For example,
\begin{equation}\label{eqn:BEFsolexpl}
 F_{3,4}=\<g^-g^-\la^-\la^+g^+g^+\>
        =\frac{\<4|(1+2)|3]\<g^-g^-g^-g^+g^+g^+\>
        +\<12\>[56]\<\la^-\la^-\la^-\la^+\la^+\la^+\>}
         {(p_1+p_2+p_3)^2} \,.
\end{equation}
A similar formula for all four-gluino two-gluon amplitudes can be 
found in the appendix of ref.~\cite{Bianchi2008pu}.


\subsection{The second $\CN=1$ SUSY diamond}\label{sec:seconddiamond}

Recall~\cite{Stieberger2007am} that the pure-gluon
amplitude can be calculated from the latter three amplitudes
in~\eqn{eqn:availamps}, namely
\begin{equation}
\<\phi^-\phi^-\phi^-\phi^+\phi^+\phi^+\>,\quad
\<\phi^-\phi^-\la^-\la^+\phi^+\phi^+\>\quad\text{and}
\quad\<\phi^-\phi^-g^-g^+\phi^+\phi^+\> \,.
\end{equation}
The question that immediately arises is whether this set
forms a basis for the complete set of all six-point NMHV
$\CN=2$ amplitudes\footnote{The term $\CN=2$ amplitudes refers to all
possible amplitudes that can be constructed exclusively from 
particles from a single $\CN=2$ multiplet and its CPT conjugate,
$(g^\pm,\la_m^\pm,\phi^\pm)$ with $m=1,2$~\cite{Sohnius1985qm}.}
in helicity configuration $X$.  We were not aware of a direct answer
to that question, so we took the following approach. As mentioned already
in subsection~\ref{sec:suitamp}, we will consider a second set of
six-point NMHV $\CN=1$ supersymmetric Ward identities,
in addition to the $\CN=1$ SWI for $(g^\pm,\la^\pm)$
described in the previous subsection.

\begin{figure}[h]
\psset{xunit=9pt,yunit=5pt,runit=2.5pt,linewidth=1pt}
\begin{pspicture}(-20,0)(28,24)
 \psdiamond[linecolor=gray,linewidth=1pt,dimen=inner](0,12)(12.125,12.125)
 \psdiamond[fillstyle=solid,fillcolor=lightgray,linecolor=darkgray,
linewidth=0.5pt,dimen=middle](0,6)(6,6)
 \psdiamond[fillstyle=solid,fillcolor=lightgray,linecolor=darkgray,
linewidth=0.5pt,dimen=middle](0,18)(6,6)
 \psdot[dotsize=3.5pt,linecolor=black](0,24)
 \psdot[dotsize=3.5pt,linecolor=black](0,12)
 \psdot[dotsize=3.5pt,linecolor=black](0,0)
 \multips(-4,20)(1,0){9}{\psdot[dotsize=3.5pt,linecolor=black](0,0)}
 \multips(-4,16)(1,0){9}{\psdot[dotsize=3.5pt,linecolor=black](0,0)}
 \multips(-4,8)(1,0){9}{\psdot[dotsize=3.5pt,linecolor=black](0,0)}
 \multips(-4,4)(1,0){9}{\psdot[dotsize=3.5pt,linecolor=black](0,0)}
 \pscircle[linecolor=black,linewidth=1.3pt](0,24){2}
 \pscircle[linecolor=black,linewidth=1.3pt](0,0){2}
 \pscircle[linecolor=black,linewidth=1.3pt](-2,4){2}
 \psframe[linecolor=black,linewidth=1.3pt](-0.6,11)(0.6,13)
 \rput[l](-14.5,8){\textcolor{black}{$\CN=2$}}
 \rput[l](-10,4){\textcolor{black}{$\CN=1$}}
 \rput[l](-10,20){\textcolor{black}{$\CN=1$}}
 \psarc[linecolor=darkgray,linewidth=0.6pt]{-}(-6.5,8){20pt}{270}{320}
 \psarc[linecolor=darkgray,linewidth=0.6pt]{-}(-6.5,16){20pt}{40}{90}
 \psarc[linecolor=darkgray,linewidth=0.6pt]{-}(-10.9,12){20pt}{270}{313}
 \rput[l](14,24){\textcolor{black}{$\<g^-g^-g^-g^+g^+g^+\>$}}
 \rput[l](14,20){\textcolor{black}{$\<g^-g^-\la^-\la^+g^+g^+\>$}}
 \rput[l](14,16){\textcolor{black}{$\<g^-\la^-\la^-\la^+\la^+g^+\>$}}
 \rput[l](14,12){\textcolor{black}{$\<\la^-\la^-\la^-\la^+\la^+\la^+\>$}}
 \rput[l](14,8){\textcolor{black}{$\<\phi^-\la^-\la^-\la^+\la^+\phi^+\>$}}
 \rput[l](14,4){\textcolor{black}{$\<\phi^-\phi^-\la^-\la^+\phi^+\phi^+\>$}}
 \rput[l](14,0){\textcolor{black}{$\<\phi^-\phi^-\phi^-\phi^+\phi^+\phi^+\>$}}
\end{pspicture}
\caption{Amplitudes involving particles from a single $\CN=2$ multiplet
containing two $\CN=1$ subsets.}
\label{fig:N2}
\end{figure}
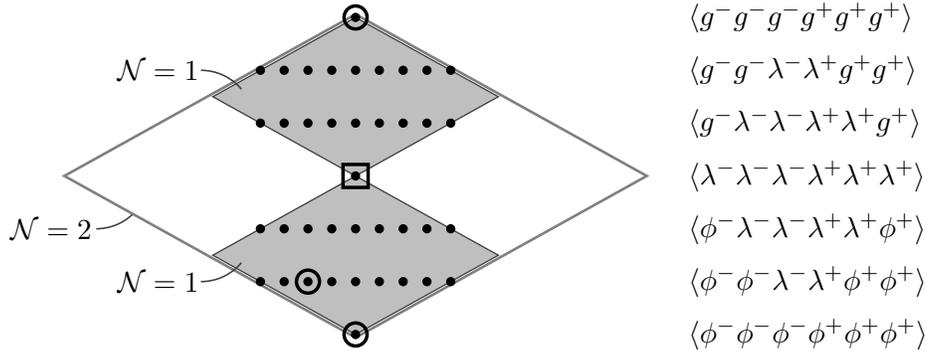

In figure~\ref{fig:N2} the collection of six-point NMHV
$\CN=2$ amplitudes is depicted in helicity configuration $X$. Every
black dot denotes a particular amplitude.  The top point
represents the pure-gluon amplitude $\<g^-g^-g^-g^+g^+g^+\>$, the
lowest point refers to the pure-scalar amplitude
$\<\phi^-\phi^-\phi^-\phi^+\phi^+\phi^+\>$, and the central point
denotes the pure-gluino amplitude
$\<\la^-\la^-\la^-\la^+\la^+\la^+\>$. Supersymmetric Ward identities
relate certain amplitudes from adjacent rows and the elements of
\eqn{eqn:availamps} are encircled. The upper diamond-shaped region
corresponds precisely to figure~\ref{fig:N1SWI}: it is the subset of
six-point NMHV $\CN=1$ amplitudes built from the multiplet
$(g^\pm,\,\la^\pm)$ within the $\CN=2$ amplitudes.
(There are additional states in the full $\CN=2$ diamond in
figure~\ref{fig:N2}, of course, even in the second row.)

However, the upper diamond-shaped region is not the only subset of
six-point NMHV $\CN=2$ amplitudes which can be related by $\CN=1$
supersymmetric Ward identities.  Stretching between the pure-gluino and
the pure-scalar amplitude there is a second region (referred to as the
lower diamond in the following), which satisfies relations similar to
those in the upper $\CN=1$ diamond. The modified supersymmetry
operator $\tilde Q$ will now act on a multiplet consisting of scalars
$(\phi^+,\phi^-)$ and gluinos $(\la^+,\la^-)$ via
\begin{align}
\left[\tilde Q(\eta),\phi^+(p)\right]&=\<p\eta\>\la^+(p),\nonumber\\
\left[\tilde Q(\eta),\la^+(p)\right]&=-\left[p\eta\right]\phi^+(p),\nonumber\\
\left[\tilde Q(\eta),\phi^-(p)\right]&=\left[p\eta\right]\la^-(p),\nonumber\\
\left[\tilde Q(\eta),\la^-(p)\right]&=-\<p\eta\>\phi^-(p)\,,
\end{align}
which can be easily derived by identifying the supercharges of $\CN=2$
supersymmetry, $Q_1$ and $Q_2$, with $Q$ and $\tilde Q$
respectively. 

Writing down the set of supersymmetric Ward identities
generated by acting with a supersymmetry generator $\tilde Q$ on the
source term $\<\phi^-\phi^-\phi^-\la^+\phi^+\phi^+\>$, one encounters
the same structure derived in \eqn{eqn:pgSWI}:
\begin{align}
 [4\eta]\<\phi^-\phi^-\phi^-\phi^+\phi^+\phi^+\>-[1\eta]
\<\la^-\phi^-\phi^-\la^+\phi^+\phi^+\>&\nonumber\\
\null -[2\eta]\<\phi^-\la^-\phi^-\la^+\phi^+\phi^+\>-[3\eta]
\<\phi^-\phi^-\la^-\la^+\phi^+\phi^+\>&=0.
\end{align}
In fact, one can show that the complete system of supersymmetric Ward
identities and amplitudes for the lower diamond, ranging from the
pure-gluino to the pure-scalar amplitude, can be obtained from the original
$\CN=1$ system considered in figure~\ref{fig:N1SWI} by exchanging
\begin{align}\label{eqn:exchange}
 Q\quad&\leftrightarrow\quad\tilde{Q} \nonumber\\
 \left[\quad\right]\quad&\leftrightarrow\quad\<\quad\> \nonumber\\
g^+\quad&\leftrightarrow\quad\phi^+ \nonumber\\
g^-\quad&\leftrightarrow\quad\phi^-.
\end{align}
This symmetry corresponds geometrically to reflecting
figure~\ref{fig:N2} about a horizontal line passing through the
central point $\<\la^-\la^-\la^-\la^+\la^+\la^+\>$.

The second system of supersymmetric Ward identities in the lower diamond is
obviously of the same rank as the original system. However, in contrast
to the upper diamond it contains two of the known amplitudes from
ref.~\cite{Stieberger2007am},
\begin{equation}\label{eqn:avampslower}
\<\phi^-\phi^-\phi^-\phi^+\phi^+\phi^+\>
\and\<\phi^-\phi^-\la^-\la^+\phi^+\phi^+\>\,,
\end{equation}
which allows the calculation of any other amplitude in the
lower $\CN=1$ set. In particular, the pure-gluino amplitude
$\<\la^-\la^-\la^-\la^+\la^+\la^+\>$
(\psset{xunit=5pt,yunit=5pt,runit=2.5pt,linewidth=1.5pt}
\begin{pspicture}(0,0)(2,2)\psframe(0.1,-0.4)(1.9,1.4)
\end{pspicture}
in figure~\ref{fig:N2}), which is the element connecting the upper and
lower set of equations, can be determined. Having done so, there are
now two known amplitudes from the upper $\CN=1$ diamond, the
pure-gluino and the pure-gluon amplitude~\cite{Stieberger2007am},
which in turn is the precondition for determining any amplitude from
the upper $\CN=1$ region. In other words: any six-point NMHV amplitude
in the two shaded regions in figure~\ref{fig:N2} can be calculated
from \eqn{eqn:availamps}.

In the next section, we will complete the ellipses on the left-hand side
of the decomposition~\eqref{eqn:choice5} by two gravitini and two gravitons,
and KLT-factorize the result in such a way that the desired six-point
closed-string ($\CN=8$ supergravity) amplitude can be related to a set of
two-gluino four-gluon $\CN=4$ SYM amplitudes.  The SYM amplitudes are
available in turn by the two-step procedure described above.


\section{$E_{7(7)}$ symmetry for $\a'$-corrected amplitudes?}
\label{sec:calculation}

As explained in the last section, the most accessible way of testing
the double-soft scalar limit relation is to calculate the $\CN=8$
supergravity amplitude,
\begin{equation}\label{eqn:DSL}
\<X^{1234}\,X_{1235}\,F^{5+}F_4^-\,B^+\,B^-\>
=\text{KLT}\Bigl[\<g^-\,\la^{4+}\,g^+\,\la_4^-\,g^+\,g^-\>_L
\times\<g^+\,\la_5^-\,\la^{5+}\,g^-\,g^+\,g^-\>_R\Bigr]\,,
\end{equation}
a particular version of \eqn{eqn:choice5}.  The determination of the
right-hand side of \eqn{eqn:DSL} will be done by employing the
two-step procedure described in the last subsection.

How should we obtain the pure-gluino amplitude
$\<\la^-\la^-\la^-\la^+\la^+\la^+\>$ from the amplitudes
in~\eqn{eqn:avampslower} in the first step?
An expression relating any six-point NMHV two-fermion
four-boson amplitude to the pure-fermion and pure-boson one has been
given in \eqn{eqn:BEFsol}.  We start from \eqn{eqn:BEFsolexpl},
employ the correspondence~(\ref{eqn:exchange}) which transforms the 
pure-gluon amplitude into the pure-scalar one, and solve the resulting
equation for the pure-gluino amplitude:
\begin{equation}
 \<\la^-\la^-\la^-\la^+\la^+\la^+\>=\frac{(p_1+p_2+p_3)^2
\<\phi^-\phi^-\la^-\la^+\phi^+\phi^+\>-\<3|(1+2)|4]
\<\phi^-\phi^-\phi^-\phi^+\phi^+\phi^+\>}{\<56\>[12]}\,.
\end{equation}
In the second step, we employ \eqn{eqn:BEFsol} to obtain
analytical expressions for all two-gluino four-gluon amplitudes,
allowing us to assemble finally the $\CN=8$ amplitude.

In the same manner as explained in subsection~\ref{sec:KLT} for
the expansion to $\CO(\a'^2)$ of a five-point gravity amplitude,
appropriate combinations of orders in $\a'$ have to be added and
permuted on the right-hand side of~\eqn{eqn:DSL}
in order to obtain the result including the $\CR^4$
perturbation.  Explicitly, the third order in $\a'$ can be
obtained by evaluating
\begin{align}
 M_6^{\CO(\a'^3)} &= \frac{-i}{\a'^3}
 s_{12} s_{45} \biggl(
 A_6^\SYM(1,2,3,4,5,6) \nonumber\\
&\hskip2.3cm
\times \Bigl[ s_{35} A_6^{\CO(\a'^3)}(2,1,5,3,4,6)
+ (s_{34}+s_{35}) A_6^{\CO(\a'^3)}(2,1,5,4,3,6) \Bigr] \nonumber\\
&\hskip2cm \null
+ A_6^{\CO(\a'^3)}(1,2,3,4,5,6) \nonumber\\
&\hskip2.3cm
\times \Bigl[ s_{35} A_6^\SYM(2,1,5,3,4,6)
+ (s_{34}+s_{35}) A_6^\SYM(2,1,5,4,3,6) \Bigr]
 \biggr) \nonumber\\
& \quad\quad\qquad\qquad + \CP(2,3,4)\,.
\label{sixptalphap3}
\end{align}
All amplitudes needed on the right-hand side of \eqn{sixptalphap3}
are two-gluino four-gluon amplitudes for the helicity configurations
$X$, $Y$ or $Z$, which we have related by supersymmetry to the
amplitudes considered in ref.~\cite{Stieberger2007am}.

Before discussing the double-soft limit relation, we examine the
single-soft limits, testing to see whether the
vanishing~(\ref{eqn:singlesoft}) observed in $\CN=8$ supergravity still
holds for the $\CR^4$ matrix elements.  For the four-point
amplitude, the factor of $s_1s_2(s_1+s_2)$ in the $\CO(\a'^3)$ term in 
$V^{(4)}_\closed$ in \eqn{eqn:closedstringexpansion}
shows that the $\CR^4$ matrix element vanishes at least as fast
as the supergravity amplitude.  Similarly, using the
forms~(\ref{56MHVopen}) for the open string five- and six-point MHV
amplitudes, together with the appropriate KLT relations, we find
numerically that the single-soft limit of the five- and six-point MHV
matrix elements of $\CR^4$ vanish.  That is, we construct a sequence
of kinematical configurations with the momentum of the scalar tending
to zero, and we find that the $\CR^4$ matrix elements vanish.
The vanishing is at the same rate as for the supergravity amplitudes,
linearly in the soft scalar momentum.  (In the MHV case, it is sufficient to
test the single-soft vanishing for one particular amplitude containing scalars,
because all other MHV amplitudes are related by SWI involving ratios of
spinor products that are constant in the soft limit.)

On the other hand, when we examine the single-soft limit of the non-MHV
six-point $\CR^4$ matrix element~(\ref{sixptalphap3}) numerically, we
find that it does {\it not} vanish.\footnote{We thank Juan Maldacena for
  suggesting that we examine this limit, and for related discussions.}
The question is whether this implies the breaking of $E_{7(7)}$ symmetry
by the $\CR^4$ term.  In principle there could be modifications to
the external scalar emission graphs that still allowed the symmetry to be
intact (as happens in the pion case).  However, the $\CR^4$ term does not
produce any nonvanishing on-shell three-point amplitudes.  So it seems
that the $E_{7(7)}$ symmetry is indeed broken, beginning at the level of
the non-MHV six-point amplitude.

One might wonder why the breaking shows up only at
this level.  If we consider the ten-dimensional term
$e^{-2\phi}t_8t_8 R^4$ discussed in the introduction, which becomes 
$e^{-6\phi}t_8t_8 R^4$ after transforming to Einstein frame, one might
suspect a violation of
the single-soft limit from the non-derivative $\phi$ coupling already at
the five-point level, expanding $e^{-6\phi}=1-6\phi+\ldots$, and
with $R^4$ producing two negative and two positive
helicity gravitons.  However, in four dimensions, the dilaton belongs to
the $\mathbf{70}$ of $SU(8)$, while the gravitons are singlets, so a
$\langle \phi B^- B^- B^+ B^+ \rangle$ amplitude is forbidden by $SU(8)$.
Adding another scalar corresponds to providing a quadratic $SU(8)$-invariant 
scalar prefactor for $R^4$, and first affects NMHV six-point
amplitudes.

Despite the apparent breaking of the $E_{7(7)}$ symmetry exhibited by
the single-soft limit of the NMHV six-point amplitude
$\<X^{1234}\,X_{1235}\,F^{5+}F_4^-\,B^+\,B^-\>$ at $\CO(\a'^3)$,
we now proceed to examine the double-soft limits of this amplitude.
First, though, we turn to the right-hand side of the double-soft limit
relation~\eqref{eqn:ACK}.  Given the particular choice of
amplitude~\eqref{eqn:DSL}, it is straightforward to find an expression for
the right-hand side. The operator 
\begin{equation}
T^4_{\,\,5}
=\ve^{12345}_{12354}\,\,\eta_{i5}\pd_{\eta_{i4}}
=-\,\eta_{i5}\pd_{\eta_{i4}}
\end{equation}
will act on the remnant of the six-point amplitude as
 \begin{align}
 &-\sum_{i=3}^{6}\eta_{i5}\pd_{\eta_{i4}}
 \<F^{5+}F_4^-\,B^+\,B^-\>\nonumber\\
 &=\sum_{i=3}^{6}\eta_{i5}\pd_{\eta_{i4}}
 \left(\frac{\pd}{\pd\eta_{35}}\right)
 \left(\frac{1}{7!}\ve_{12345678}
 \frac{\pd^7}{\pd\eta_{41}\ldots\pd\eta_{43}\pd\eta_{45}\ldots\pd\eta_{48}}\right)
 \left(\frac{1}{8!}\ve_{12345678}
 \frac{\pd^8}{\pd\eta_{61}\ldots\pd\eta_{68}}\right)\Omega_4\nonumber\\
   &=\<F^{4+}F_4^-\,B^+\,B^-\>-\<F^{5+}F_5^-\,B^+\,B^-\> \,.
\label{RHSaction}
\end{align}
Acting on particle $3$, the operator changes the derivative 
with respect to $\eta_{35}$ into a derivative with respect to $\eta_{34}$, 
thus effectively transforming the positive helicity gravitino $F^{5+}$ 
into $F^{4+}$.  Correspondingly, by acting on particle $4$, again a 
derivative with respect to $\eta_{45}$ will be changed into one with
respect to $\eta_{44}$, this time transforming $F_4^-$ into $F_5^-$. 

Restoring the kinematical weight factors in \eqn{eqn:ACK}, 
the final comparison will be made according to the following formula:
\begin{align}\label{eqn:finalcomparison}
 &\<X^{1234}\,X_{1235}\,F^{5+}F_4^-\,B^+\,B^-\>\Bigl|_{\CO(\a'^3)}
\ \xrightarrow\nonumber\\
 &\frac{1}{2}\left[\frac{p_3\cdot(p_2-p_1)}{p_3\cdot(p_1+p_2)}
  \<F^{4+}F_4^-\,B^+\,B^-\>\Bigl|_{\CO(\a'^3)}
                  - \, \frac{p_4\cdot(p_2-p_1)}{p_4\cdot(p_1+p_2)}
  \<F^{5+}F_5^-\,B^+\,B^-\>\Bigl|_{\CO(\a'^3)}\right] \,.
\end{align}
Given the complexity of the higher-order $\a'$ corrections in the
available amplitudes (see {\it e.g.} \eqn{eqn:AmpX} at only $\CO(\a'^2)$),
the analytical computation of the left-hand side of
\eqn{eqn:finalcomparison} would be very cumbersome.
Instead the computation and comparison have been
performed numerically for a sufficient number of kinematical points.

For reference, we give numerical values at one sample double-soft
kinematical point, with all outgoing momenta
fulfilling $p_i^2=0$ and $\sum_{i=1}^6 p_i^\mu=0$:
\begin{align}
p_1&=\ (-0.853702542142,\ +0.696134406758,\ -0.306157335124,\ +0.387907984368)
         \times 10^{-4}, \nonumber\\
p_2&=\ (+0.711159367201,\ -0.099704627834,\ -0.295472686856,\ +0.639142021830)
         \times 10^{-4}, \nonumber\\
p_3&=\ (+0.818866370407,\ +0.408234512914,\ -0.661447772542,\ -0.257630664418),
\nonumber\\
p_4&=\ (-1.098195656456,\ -0.551965696904,\ -0.598319787466,\ +0.737143813124),
\nonumber\\
p_5&=\ (-0.618073260483,\ +0.143671541012,\ +0.362410922160,\ -0.479615853707),
\nonumber\\
p_6&=\ (+0.897416800850,\ +0.000000000000,\ +0.897416800850,\ +0.000000000000).
~~~\label{PhaseSpacePoint}
\end{align}
At this point, with a particular external-state phase convention,
the left- and right-hand sides of the supergravity
($\CO(\a'^0)$) version of \eqn{eqn:finalcomparison} are given
respectively by
\begin{equation}
-0.30572232 \, - \, i\, 0.89270274
\ \approx\ 
-0.30615989 \, - \, i\, 0.89271337 \,,
\label{ap0numerics}
\end{equation}
while the desired $\CO(\a'^3)$ terms in \eqn{eqn:finalcomparison} are,
\begin{equation}
3.08397954 \, + \, i \, 9.00278816
\ \approx\ 
3.08775134 \, + \, i \, 9.00339016 \,.
\label{ap3numerics}
\end{equation}
The difference between the left- and right-hand sides is
due merely to the finite separation of the point~\eqref{PhaseSpacePoint}
from the double-soft limit.  It can be made as small as desired
by working closer to the limit, using higher precision kinematics
to avoid roundoff error.

The result is surprising: for any double-soft kinematical
configuration considered, the left- and the right-hand side of
\eqn{eqn:finalcomparison} show complete agreement within
numerical errors.

Given the available amplitudes from the two shaded regions in
figure~\ref{fig:N2}, one can perform further tests for other $\CN=8$
amplitudes.  In addition to \eqn{eqn:finalcomparison}, we have tested
the double-soft scalar limit for the following amplitudes
\begin{align}
&\<X^{1234}\,X_{1235}\,F^{5+}\,F_4^-\,F^{4+}\,F_4^-\>\Bigl|_{\CO(\a'^3)}
\ \xrightarrow\nonumber\\
 &\frac{1}{2}\left[+\,\frac{p_3\cdot(p_2-p_1)}{p_3\cdot(p_1+p_2)}
\<F^{4+}F_4^-F^{4+}F_4^-\>\Bigl|_{\CO(\a'^3)}\right.\nonumber\\
 &\hskip0.5cm\left.-\,\frac{p_4\cdot(p_2-p_1)}{p_4\cdot(p_1+p_2)}
\<F^{5+}F_5^-F^{4+}F_4^-\>\Bigl|_{\CO(\a'^3)}\right.\nonumber\\
 &\hskip0.5cm\left.-\,\frac{p_6\cdot(p_2-p_1)}{p_6\cdot(p_1+p_2)}
\<F^{5+}F_4^-F^{4+}F_5^-\>\Bigl|_{\CO(\a'^3)}
 \right]
\label{test2}
\end{align}
and
\begin{align}
 &\<X^{1234}\,X_{1235}\,X^{1235}\,X_{1235}\,X^{1235}\,X_{1234}\>
\Bigl|_{\CO(\a'^3)}\ \xrightarrow\nonumber\\
 &\frac{1}{2}\left[+\,\frac{p_3\cdot(p_2-p_1)}{p_3\cdot(p_1+p_2)}
\<X^{1234}\,X_{1235}\,X^{1235}\,X_{1234}\>\Bigl|_{\CO(\a'^3)}
\right.
\nonumber\\
 &\hskip0.5cm\left.+\,\frac{p_5\cdot(p_2-p_1)}{p_5\cdot(p_1+p_2)}
\<X^{1235}\,X_{1235}\,X^{1234}\,X_{1234}\>\Bigl|_{\CO(\a'^3)}\right.
\nonumber\\
 &\hskip0.5cm\left.-\,\frac{p_6\cdot(p_2-p_1)}{p_6\cdot(p_1+p_2)}
\<X^{1235}\,X_{1235}\,X^{1235}\,X_{1235}\>\Bigl|_{\CO(\a'^3)}
 \right]\,.
\label{test3}
\end{align}
Each limit shows complete agreement for any double-soft
kinematical point.


\section{Conclusion}
\label{sec:conclusion}

Our computation shows that the double-soft limit of three distinct
six-point $\CO(\a'^3)$-corrected $\CN=8$ matrix elements yields the
corresponding weighted sum of four-point amplitudes, precisely as 
dictated by $E_{7(7)}$ invariance~\cite{ArkaniHamed2008gz}.  
However, this is quite puzzling, given the nonvanishing single-soft limits
of the same six-point amplitudes.  The most likely possibility seems to
be that the double-soft limits will begin to fail, but only beginning
with the NMHV seven-point amplitudes.  It would be very interesting to
test this limit, but that is beyond the scope of the present paper.

Whether the three-loop cancellations~\cite{Bern2007hh,Bern2008pv} 
can be explained by a simple
symmetry argument that originates in the $\sfrac{E_{7(7)}}{SU(8)}$
coset symmetry of $\CN=8$ supergravity is still open.  This work suggests
that the $\CR^4$ term produced by tree-level string theory can be ruled
out in this way, but other dependences on scalars should be considered.
The work of Green and Sethi~\cite{Green1998by} in ten dimensions
indicates that supersymmetry may forbid any $\CR^4$ term, but an argument
using supersymmetry directly in four dimensions would be very welcome.

Of course there are higher-dimension potential counterterms than
$\CR^4$, which are relevant beginning at five loops.  It is possible 
that $E_{7(7)}$ and/or
supersymmetry can be used to exclude these counterterms as well, 
up to a certain dimension or loop order.
However, at eight loops a counterterm exists 
that is invariant under both supersymmetry and
$E_{7(7)}$~\cite{Howe1980th,Kallosh1981vz}.
It is still possible that $E_{7(7)}$ plays a more subtle
role in the excellent ultraviolet behavior of the theory, perhaps by
relating somehow the coefficients of certain loop integrals
making up the full multi-loop amplitude.

Completely understanding the role of $E_{7(7)}$ will very likely
be part of a fundamental explanation of the conjectured finiteness
of $\CN=8$ supergravity. However, whether supersymmetry and the coset 
symmetry alone are sufficient ingredients remains to be shown.


\section*{Acknowledgments}

The authors would like to express their gratitude to Stephan
Stieberger for conversations, and to Tomasz Taylor and Stephan
Stieberger for providing complete Mathematica expressions for the NMHV
six-point open-string amplitudes, which served as a starting point for
this calculation.  Furthermore we would like to thank Zvi Bern,
Michael Douglas, Henriette Elvang, Dan Freedman, Renata Kallosh,
Jared Kaplan, Hermann Nicolai, Stefan Theisen and especially
Juan Maldacena for helpful discussions.
This research was supported by the U.S. Department of
Energy under contract DE-AC02-76SF00515.  The work of J.B. was
supported by the German-Israeli Project cooperation (DIP) and the
German-Israeli Foundation (GIF).


\end{document}